\documentstyle[revtex,epsf]{aps}
\begin{document}
%\draft
\preprint{Computers in Physics Compuscript}

\begin{title}
Lattice-Gas Automata Fluids on Parallel Supercomputers
\end{title}

\author{Jeffrey Yepez}
\begin{instit}
US Air Force, Phillips Laboratory, Hanscom Field, Massachusetts
\end{instit}
\author{Guy P. Seeley}
\begin{instit}
 Radex Corporation, Bedford, Massachusetts
\end{instit}
\author{Norman Margolus}
\begin{instit}
MIT Laboratory for Computer Science, Cambridge, Massachusetts
\end{instit}

\begin{abstract}
A condensed history and theoretical development of lattice-gas automata in the
Boltzmann
limit is presented.  This is provided as background to  set up the context for
understanding
the implementation of the lattice-gas method on two parallel supercomputers:
the MIT
cellular automata machine CAM-8 and the Connection Machine CM-5.  The
macroscopic
limit of two-dimensional fluids is tested by simulating the Rayleigh-B\'enard
convective
instability, Kelvin-Helmholtz shear instability, and the Von Karman vortex
shedding
instability. Performance of the two machines in terms of both site update rate
and
maximum problem size are comparable.  The CAM-8, being a low-cost desktop
machine,
demonstrates the potential of special-purpose digital hardware.
\end{abstract}
\pacs{}

%\date{November 23, 1993}

\newcommand{\ea}{\mbox{${\bf \hat e_a}$}}
\newcommand{\e}{\mbox{${\bf \hat e}$}}
\newcommand{\r}{\mbox{${\bf r}$}}
\newcommand{\q}{\mbox{${\bf q}$}}
\newcommand{\p}{\mbox{${\bf p}$}}
\newcommand{\F}{\mbox{${\bf F}$}}
\newcommand{\f}{\mbox{${f_a}$}}
\newcommand{\x}{\mbox{${\bf x}$}}
\newcommand{\X}{\mbox{${\bf X}$}}
\newcommand{\y}{\mbox{${\bf y}$}}
\newcommand{\vel}{\mbox{${\bf v}$}}

\newcommand{\ah}{\mbox{${\hat a}$}}
\newcommand{\ahd}{\mbox{${\hat a^\dagger}$}}
\newcommand{\ch}{\mbox{${\hat c}$}}
\newcommand{\chd}{\mbox{${\hat c^\dagger}$}}
\newcommand{\exh}{\mbox{${\bf \hat{F}}$}}
\newcommand{\Ap}{\mbox{${\hat A_+}$}}
\newcommand{\Am}{\mbox{${\hat A_-}$}}
\newcommand{\Rp}{\mbox{${\hat R_+}$}}
\newcommand{\Rm}{\mbox{${\hat R_-}$}}
\newcommand{\Apd}{\mbox{${\hat A^\dagger_+}$}}
\newcommand{\Amd}{\mbox{${\hat A^\dagger_-}$}}
\newcommand{\Rpd}{\mbox{${\hat R^\dagger_+}$}}
\newcommand{\Rmd}{\mbox{${\hat R^\dagger_-}$}}

\newcommand{\vac}{\mbox{$\mid 0 \rangle$}}
\newcommand{\one}{\mbox{${\bf 1}$}}
\newcommand{\mex}{\mbox{${\bf \hat{\chi}}$}}
\newcommand{\mexint}{\mbox{${\bf \hat{\chi}^{\tiny\hbox{int}}}$}}
\newcommand{\U}{\mbox{${\bf \hat{U}}$}}
\newcommand{\N}{\mbox{${\bf \hat{N}}$}}
\newcommand{\ham}{\mbox{${\bf \hat{H}}$}}

%\narrowtext
\twocolumn

\nonum\section{Introduction}

Lattice-gas automata dynamics is a discrete form of molecular dynamics.  In
molecular
dynamics one simulates a many-body system of particles with continuous
interaction
potentials \cite{rahman-oct64}. The particles have continuous positions and
momenta.  In
lattice-gas dynamics the particles' positions and momenta are discrete and
motion is
constrained to a spacetime crystallographic lattice.

One may also view lattice-gas automata dynamics as an extension of cellular
automata,
popularized in the physics community by Stephen Wolfram
\cite{wolfram-83,wolfram-84}.
An elementary treatment of the cellular automata subject is presented  by
Tommaso Toffoli
and Norman Margolus in their book on cellular automata machines
\cite{toffoli-87}.
Corresponding to  the cellular automata paradigm, lattice-gas automata are
ideally suited
for massively parallel processing.  In lattice-gas automata models, each
possible
momentum state at a given position is represented by a single digital bit.
Therefore, a
Pauli exclusion principle is enforced where there can be no more than a single
particle per
momentum state. As a particle in state $\alpha$ at some lattice site of the
crystallographic
space ``hops'' into state $\beta$, say at a neighboring site, a digital bit is
moved from
$\alpha$ and into $\beta$.  So in lattice-gas dynamics one simulates a system
of boolean
particles where the data streaming corresponds to spatial translation and the
data
permutation corresponds to collisional interactions.  This computational analog
of particle
dynamics offers an exciting alternative to, and not simply an approximation of,
the usual
partial differential equations\cite{toffoli-84a}.  So the lattice-gas
methodology has an
intrinsic value beyond finite difference schemes.

The lattice gas approach has been extended to a lattice-based
Krook-Bhatnager-Gross
approximation of the Boltzmann equation
\cite{liboff-79,rivet-86,mcnamara-88,chen-jun91}, which we call the
lattice-Boltzmann
equation. In place of the exactly computable dynamics of boolean particles, one
focuses
on a statistical regime where a particle has a probability of occupying a given
momentum
state. Moreover, many particles can occupy the same momentum state at the same
position. The approach offers both theoretical and computational advantages.
An
important theoretical advantage is that one may capture the essential physics
of the
complex system by stating no more than the system's equilibrium distribution.
Computationally it has the advantage of reducing noisy fluctuations in the
system at the
expense of discarding information concerning particle-particle correlations.
Lattice-gas
automata in the Boltzmann approximation have become one of the most important
contributions by the lattice-gas community to high-performance computational
physics
\cite{karniadakis-93}.  The lattice-Boltzmann equation, usually implemented
within one
hundred lines of code on a massively parallel processor \footnotemark[1],
allows the
researcher to efficiently model complex systems, providing a straightforward
particle-based metaphor to computation. Yet it relies on expensive floating
point
calculations and therefore is most suited for massively parallel machines such
as the
Connection Machine-5 (CM-5).  In this paper, we use the lattice-Boltzmann
equation only
for theoretical analysis---all of our simulations are based on lattice gases.

We have implemented a general lattice-gas automaton on two parallel
architectures: the
experimental MIT CAM-8 and the Thinking Machines Corporation CM-5.  Very
briefly, a
CAM-8 node has $2^{22}$ 16-bit sites (8 Mbytes DRAM) and a double buffered
look-up
table (2 Mbits SRAM), with a clock speed of 25 MHz.  There are no processors,
only
lookup tables.  The nodes are connected in a 3D mesh. A CM-5 node has 32 Mbytes
of
DRAM, four vector units at 16 MHz , and a SPARC processor at 32 MHz. Its nodes
are
connected by a fat-tree network.  The architectures of both machines will be
discussed in
more detail below.

The presentation here is limited to two-dimensional fluids. Three-dimensional
hydrodynamics, immiscible fluids
\cite{rothman-89,chen-91,gustensen-91a,gustensen-91b}, multiphase systems
\cite{chen-sep89,appert-90,appert-91,yepez-93}, reaction-diffusion systems,
magnetohydrodynamics \cite{chen2}, and flow through porous media
\cite{rothmann-90,chen-91b} are also subjects of active research.  We have
implemented
hydrodynamic and thermohydrodynamic lattice-gases on the CAM-8, the first
lattice-gas
experiment conducted on the prototype machine.  Results of the CAM-8 and CM-5
simulations of the Rayleigh-B\'enard convection instability, Kelvin-Helmholtz
shear
instability, and Von-Karman vortex shedding instability are presented.

Our main findings are the following.  The CAM-8 delivers 25 million site
updates per
second per module for a 16 bit lattice gas. The CM-5 in principle can simulate
larger
lattices due to its much larger memory size (16Gbytes on a 512-node partition).
In practice
the typical problem sizes were almost identical on a 128-node partition of the
CM-5 and on
an 8-module CAM-8 prototype: $4096\times2048$ 2D lattices on both machines.
The
CM-5 can deliver about 1 million site updates per second per node for a 16 bit
lattice gas.
For the CM-5 interprocessor communication cost is small relative to the
computation
involved in the data streaming and site update when the size of the lattice is
large enough.
This suggests that the relatively low communication bandwidth imposes no
serious
degradation on the delivered performance.  We also find that the delivered
performance
increases as the lattice size increases, see appendix~\ref{c*}.

The best way to report performance of each machine for lattice-gas simulation
is to give
the site update rate.  The best rate achieved on the thermohydrodynamic
lattice-gas was
191 million site updates per second on an 8-module CAM-8 and 110 million on on
a
256-node partition of the CM-5.  For a simpler problem, the FHP lattice-gas, a
256-node
partition of the CM-5 attained a site update rate of 550 million updates per
second on a
32K by 2K lattice.  The best rate obtainable on the 8-module CAM-8 for the same
simple
problem is 382 million site updates per second.  The CAM-8 prototype is a
desktop
machine (see figure~\ref{cam-8_prototype}) containing approximately the same
amount of
digital logic and memory as is found in a common workstation.  The efficiency
of the
CAM-8 comes at a severe price, in terms of its specialization to a certain
class of scientific
computations, albeit its applicability is widening \cite{margolus-93}.

\section{Some Historical Notes}

Let us briefly review some of the historical developments of the lattice-gas
subject.   An
overview of the lattice-gas subject has also been given by Boghosian
\cite{boghosian-91}.
Simple implementations of a discrete molecular dynamics on a square lattice
were
investigated in the early 1970's by the French, in particular Yves Pomeau and
coworkers
\cite{hardy-76}. By the late 1970's, cellular automata research was underway at
the
Information Mechanics Group at MIT on reversible computation by Edward Fredkin,
Tommaso Toffoli, and Norman Margolus\cite{toffoli-77b,fredkin-82,margolus-84}.
The idea
of building special-purpose machines to simulate physics-like models on a
fine-grained
space \cite{toffoli-84b,margolus-84} originated there and  today still remains
a strength of
that group. A good review of the kind of cellular automata modeling done in the
early
1980's is given by G\'erard Vichniac \cite{vichniac-84}. During this time,
Stephen Wolfram
visited the Information Mechanics Group and was stimulated by their work. In
1983
Wolfram popularized cellular automata as a simple mathematical model to
investigate
self-organization in statistical mechanics\cite{wolfram-83,packard-85}.

After visiting the MIT Information Mechanics Group in 1983 and seeing a TM-gas
simulation on the CAM-5 machine of Toffoli and
Margolus\cite{toffoli-84a,toffoli-84b},
Pomeau realized the potential for simulating large fluid systems and much new
interest and
activity in the field emerged. A race began to theoretically prove that a
hydrodynamic limit
emerges from simple lattice-gas automata.  The intense interest was not stirred
as much by
the subject of hydrodynamics itself, but instead by the possibility of a simple
cellular
space-time model capturing such complex natural behavior in an exact way. In
1985
Wolfram completed the first hydrodynamics simulations on a triangular lattice
\cite{Wolfram-85} on the Connection Machine---at this time, lattice-gases were
one of the
most important applications for the bit oriented single instruction multiple
data Connection
Machine. By 1986  Frisch, Hasslacher, and Pomeau had reported the existence of
an
isotropic two-dimensional lattice-gas on the triangular lattice
\cite{frisch-86}.  In the same
year Wolfram completed the most detailed treatment of the basic theory
including novel
symmetry considerations and introduced the Boltzmann approximation. Frisch {\it
et al.}
found the minimal lattice symmetry needed to recover isotropic flow in the
continuum limit
is a triangular lattice with a particle possessing six momentum states. Their
model is now
referred to as the FHP-model or hexagonal lattice-gas model. Accompanying the
seminal
1986 FHP paper was a paper by Margolus, Toffoli, and Vichniac on
cellular-automata
supercomputers for fluid-dynamics modeling \cite{margolus-86}.  The
contribution of
Margolus {\it et al.}  complements the theoretical work of Frisch {\it et al.},
pointing out that
with dedicated computational hardware the lattice-gas model potentially gains a
unique
advantage over traditional methods of physical modeling.

By 1987 the lattice-gas methodology was extended to model three-dimensional
flows.  The
minimal lattice found by Frisch et el. \cite{frisch-87} was the face centered
hypercubic
(fchc) lattice.  The fchc lattice with 24-nearest neighbors is projected onto
three
dimensions in a simple fashion by limiting the depth of the fourth dimension of
the
simulation volume to one lattice link. Research is still underway on finding
optimal
collisions to minimize the viscosity of the fluid \cite{henon-90}, however this
task has
proven very difficult.  The reason for this difficulty is that the fchc
lattice-gas has $2^{24}$
or 16.7 million input configurations.  In practice, all possible collisions are
not included in a
simulation because of the large demand for local memory needed to pre-store all
the
necessary collisional events in table look-up format---an efficient format for
implementing
complex interactions.  To ease memory loads, lattice isometries are exploited
to reduce
the size of look-up tables \cite{somers-89}.

The hope of modeling very high Reynolds number flows by lattice-gas automata
methods
has not yet been realized with models that do not violate semi-detailed
balance.  However,
lattice-gas models in the Boltzmann approximation have shown considerably more
success in achieving high Reynolds number flows\footnotemark[2].

Recently the first prototype of the next generation cellular automata machine,
CAM-8, has
been constructed \cite{margolus-90b}. The current 8-module CAM-8 prototype,
with a site
density of $2^{22}$ 16-bit sites per module, has a total of 32 million sites.
Within the next
few years a large CAM-8 sponsored by the US Air Force will be constructed with
at least
$10^9$ sites and will have a computational rate of approximately 12.5 billion
site updates
per second.  This site update rate is about two orders of magnitude faster than
that
achievable with current parallel computers such as the Connection Machine,
CM-5.  In
\S\ref{gallery} we present some simulation results on the CAM-8 prototype and
the CM-5.

\section{Why lattice-gases?}

There are many reasons for studying lattice-gases, both practical and
theoretical.  Some
commonly cited reasons are oriented towards computer science and issues related
to
massively parallel processing. There are also appealing reasons related to
modeling
physical systems with complex boundary conditions. The lattice-gas' attributes
include: 1)
bit efficiency; 2) inherent simplicity; 3)logic density; and 4) exact
computability.

Firstly, lattice-gas automata allow for high bit efficiency. A single digital
bit is used to
represent a particle.  Unlike in floating point calculations where there exist
uncontrolled
round-off errors in the least significant bits, in lattice-gases all bits have
equal weight, or to
quote Frisch, there is ``bit democracy.''  Consequently, the efficiency with
which bits are
used may be higher for lattice-gases.

Secondly, lattice-gas automata possess an inherent simplicity.  Just as simple
models in
statistical mechanics, such as the Ising model, shed light on equilibrium
critical
phenomena, so too do lattice-gas models shed light on dynamical
phenomena\cite{yepez-93}. Moreover, their inherent simplicity gives them
pedagogical
value since many properties of macroscopic systems can be understood through
analytical expressions given very simple local rules.  For example,
lattice-gases are a
simple way to understand details of fluid systems such as the dependence of the
shear
viscosity on particle collision rates. Computational fluid dynamics codes are
complicated
and intricate in their approximations.  Lattice-gases are perhaps the simplest
expression of
Navier-Stokes flows and are easily implemented.

Thirdly, the combination of bit efficiency with the simplicity and locality of
some lattice-gas
rules allows---in principle---nearly ideal logic density.  At the highest logic
density that is
physically possible, there is the interesting prospect of lattice-gas
architectures built out of
``quantum hardware.''  There is the expectation that in the future, computation
will be
achieved on quantum computers
\cite{margolus-90a,lloyd-93,biafore-94,heitmann-93}.  As
the fundamental computational element's size reduces to nano-scale ranges its
behavior is
governed by quantum mechanics.  Quantum mechanics requires unitary, and hence
invertible, time evolution---the microscopic reversibility of the lattice-gas
dynamics is
important here.  Even before quantum mechanics becomes a constraint, the
reversibility of
lattice gas dynamics may become a significant benefit, since at very high logic
densities
the dissipation of heat caused by irreversible computations will become an
issue\cite{bennett-79,bennett-82}.

Fourthly, lattice-gas automata are exactly computable. Richard Feynman
\cite{feynman-82} considered on a discrete spacetime lattice ``the possibility
that there is
to be an {\it exact} simulation, that the computer will do {\it exactly} the
same as nature'',
and that using computers in an exactly computable way may lead to new
possibilities in
our understanding of physics.  Although lattice-gases cannot model quantum
systems
\cite{feynman-82}, they do model classical systems while keeping mass,
momentum, and
energy exactly conserved. Exact modeling is valuable, for example, in cases
where
multiparticle correlations are essential to the system's behavior.  Lattice-gas
simulations
can verify theoretical predictions beyond the Boltzmann mean-field
approximation of
uncorrelated collisions: the phenomenon of long-time tails in the velocity
autocorrelation
function \cite{alder-jun67,pomeau-sep68,ernst-nov70} has recently been observed
in
lattice gases   \cite{kirkpatrick-dec91,brito-dec91,brito-jul92}.

\section{Lattice-Gas Automata}
\label{lattice-gas-automata}

We first define, in the usual way, what a lattice-gas cellular automaton is.
Then we
analytically treat the lattice-gas in the Boltzmann limit to show that one may
use strictly
deterministic local rules to obtain the correct macroscopic limit.  We show in
particular that
a chiral system is adequate to obtain correct hydrodynamics. We then summarize
the
derivation of the equations of motion for a thermal lattice-gas. Finally, we
discuss our
conventions for embedding a hexagonal lattice in a square lattice. We show the
streaming
relations used here for local and nonlocal collisions.

\subsection{Some Preliminaries about the Local Dyanamics}

Variables used are the following

{\footnotesize
\begin{eqnarray*}
Mass \; Unit & : & m \\
Spatial \; Unit & : & l \\
Temporal \; Unit & : & \tau \\
Particle \; Speed  & : & c = \frac{l}{\tau} \\
Sound \; Speed  & : & c_s \\
\# \; Momentum \; Directions & : & B \\
Lattice \; Vectors & : & \e_a \\
a  & = & 1,2,\dots, B \\
Particle \; Number \; Variable & : & n_a \\
Distribution \; Function & : & f_a \\
Collision \; Operator & : & \Omega_a \\
Jacobian \; Matrix & : & J_{ab} \\
Number \; Density & : & n \\
Mass \; Density & : & \rho = m n\\
Bulk \; Velocity & : & {\bf v} \\
Total \; Internal \; Energy & : & \varepsilon
\end{eqnarray*}
}

Consider a spacetime lattice with $N$ spatial sites, unit cell size $l$, and
time unit $\tau$.
Particles, with mass $m$,  propagate on the lattice with speed $c=l/\tau$. The
lattice
vectors are denoted by $\e_a$ where $a=1,2,\dots,B$.   A particle's state is
completely
specified at some time, $t$,  by specifying its position on the lattice, $\x$,
and its
momentum, $\p=mc \ea$. The particles obey Pauli exclusion since only one
particle can
occupy a single momentum state at a time.  The total number of configurations
per site is
$2^{B}$. The total number of single particle states available in the system is
$N_{\hbox{\tiny total}} = B N$.  With $P$ particles in the system, we denote
the filling
fraction by $d=P/N_{\hbox{\tiny total}}$.

The number variable, denoted by $n_a(\x,t)$, has the value one if a particle
exists at site
$\x$ at time $t$ in momentum state $mc \ea$  and zero otherwise.   The
evolution of the
lattice-gas can be written in terms of $n_a$ as a  two-part collision and
streaming process.
The collision part permutes the particles locally at each site
\begin{equation}
n_a' (\x, t) = n_a(\x,t) + \Omega_a ( \vec{n}(\x,t) ),
\end{equation}
where $\Omega_a$ represents the collision operator and in general depends on
all the
particles at the site.  The streaming part permutes the particles globally.  A
particle at
position $\x$ ``hops" to its neighboring site at $\x+l\ea$ and then time is
incremented by
$\tau$
\begin{equation}
\label{eq:cellular-update}
n_a' (\x+l\ea, t+\tau) = n_a(\x,t) + \Omega_a ( \vec{n}(\x,t) ).
\end{equation}
Equation (\ref{eq:cellular-update}) is the lattice-gas cellular automaton
equation of motion.
Because the dynamics only permutes the occupation of states, the system is
strictly
reversible, see figure~\ref{collision_streaming}.

\subsection{Coarse-Grained Dynamics}
\label{coarse-grained-dynamics}

To simplify the theoretical analysis of the lattice-gas dynamics, it is
convenient to work in
the Boltzmann limit where a field point is obtained by a block average over the
number
variables.  That is, we may define a single particle distribution function,
$f_a = \langle n_a
\rangle$.It should be understood that whenever the single particle distribution
function is
written, its  subscripted index is taken modulo B
\begin{equation}
f_{a+b} = f_{\bmod_B (a+b)} .
\end{equation}
Using the Boltzmann molecular chaos assumption the averaged collision operator
simplifies to $\langle \Omega_a \rangle = \Omega_a (\langle \vec{n}\rangle)$,
and by a
Taylor expansion (\ref{eq:cellular-update}) we obtain the lattice Boltzmann
equation
\begin{equation}
\label{lattice-boltzmann-equation}
\partial_t f_a + c e_{ai}\partial_i f_a = \Omega_a .
\end{equation}
A careful treatment of this procedure is given by Frisch {\it et al.}
\cite{frisch-87}. A general
collision operator is constructed as follows
\begin{equation}
\Omega_a = \sum_{ \{ \zeta_i \} } \alpha Q_a ( \{\zeta_i\} ),

\end{equation}
where $\{ \zeta_i \}$ is a set of occupied particle states and $\alpha = \pm 1$
is a scalar
coefficient and where each term in the sum is written in factorized form as
\begin{equation}
Q_a ( i_1, \dots, i_k) =
\frac{f_{a+i_1}}{1-f_{a+i_1}}\cdots\frac{f_{a+i_k}}{1-f_{a+i_k}}
\prod_{j=1}^{B}(1-f_{a+j}) .
\end{equation}
We expand the distribution function about its equilibrium value,
$f^{\tiny\hbox{eq}}$
\begin{equation}
f_a = f^{\tiny\hbox{eq}} + \delta f_a

\end{equation}
so that, to first order, we have
\begin{equation}
\label{omega-expansion}
\Omega_a (f^{\tiny\hbox{eq}})  = \sum_b \frac{\partial \Omega_a}{\partial
f_{b}} \delta f_b.
\end{equation}
The l.h.s. of (\ref{omega-expansion}) must vanish, since the particle
distribution is
non-changing under equilibrium conditions. The eigenvalues of the Jacobian of
the
collision operator,

\begin{equation}
\label{boltzmann-jacobian}
J_{ab} = \frac{\partial \Omega_a}{\partial f_{b}},
\end{equation}
can be calculated and the number of these that vanish must equal the number of
invariant
quantities in the lattice-gas dynamics.  Because of the finite-point group
symmetry of the
spatial lattice, the Jacobian matrix will be circulant, its elements can be
specified by the
difference of their indices, $J_{ab} = J_{a-b}$.  This property of the Jacobian
simplifies the
solution of the eigenvalue equation
\begin{equation}
\label{eigen-equation}
\sum_b J_{a-b} \xi^k_b = \lambda^k \xi^k_a,
\end{equation}
where $k=1,\dots,B$.  Let us make the ansatz that the eigenvectors have the
following
form
\begin{equation}
\label{circulant-ansatz}
\xi^k_a = e^{2\pi i a k /B}.
\end{equation}
Then inserting (\ref{circulant-ansatz}) into  (\ref{eigen-equation}) and taking
$m=a-b$,
gives
\begin{equation}
\label{circulant-eigenvalues}
\lambda^k = \sum_m J_{m} e^{2\pi i m k /B}.
\end{equation}

\subsection{Triangular Lattice: B=6}

When implementing a lattice-gas on a parallel computer it is most convenient to
use
deterministic updating rules.  This is important for several reasons.  First of
all, using
deterministic rules, the lattice-gas possesses a strict time-reversal
invariance.  Therefore,
it mimics the time-reversal invariance of natural physical laws.  As a
practical matter, the
reversibility allows one to run the gas dynamics forward to some state and then
back to its
initial state.  This is a good way to check if the local rules are coded
correctly.  Second of
all, the generation of random numbers typically takes times and using random
bits
increases the amount of state that the rule must deal with.  Therefore,
deterministic local
rules are used here.  Lattice-gas collisions can be catergorized with even or
odd chirality.
The method employed here uses even chirality collisions on even time steps and
odd
chirality collisions on odd time steps, thereby eliminating the need for a
random coin toss.
The validity of such a partitioning of the collisions must be justified.  This
may be done in a
straight forward way using the results of \S\ref{coarse-grained-dynamics}.  The
simplest
hydrodynamic example is a definite chirality hexagonal lattice-gas, almost
identical to the
usual FHP gas except no random coin toss is made.  For a hexagonal  lattice,
$B=6$, the
eigenvectors of the Jacobian matrix, (\ref{circulant-ansatz}), are simply
composed of 1
plus the three roots of -1
\begin{eqnarray}
\xi_0 & = & (1, 1, 1, 1, 1, 1) \\
\xi_1 & = & ( \epsilon, \epsilon^\ast, -1, \epsilon, \epsilon^\ast, 1) \\
\xi_2 & = & (\epsilon^\ast, \epsilon, 1, \epsilon^\ast, \epsilon, 1) \\
\xi_3 & = & (-1, 1, -1, 1, -1, 1) \\
\xi_4 & = & (\epsilon, \epsilon^\ast, 1, \epsilon, \epsilon^\ast, 1) \\
\xi_5 & = & (\epsilon^\ast, \epsilon, -1, \epsilon^\ast, \epsilon, 1) ,
\end{eqnarray}
where $\epsilon = e^{i\frac{\pi}{3}}$.
The collision operator that produces 2-body and 3-body symmetric collision is
the
following
\begin{equation}
\Omega_{a} = Q_a (1,4) - Q_a (0,3) + Q_a (1,3,5) - Q_a (0,2,4),
\end{equation}
which is written in expanded form as
\begin{eqnarray*}
\Omega_{0} & = & -\left( f_{0}\,\left( 1 - f_{1} \right) \,\left( 1 - f_{2}
\right) \,f_{3}\,\left( 1 -

f_{4} \right) \,
     \left( 1 - f_{5} \right)  \right)  +

\\
& & \left( 1 - f_{0} \right) \,f_{1}\,\left( 1 - f_{2} \right) \,
   \left( 1 - f_{3} \right) \,f_{4}\,\left( 1 - f_{5} \right)  -

\\
& &   f_{0}\,\left( 1 - f_{1} \right) \,f_{2}\,\left( 1 - f_{3} \right)
\,f_{4}\,\left( 1 - f_{5} \right)  +

\\
& &   \left( 1 - f_{0} \right) \,f_{1}\,\left( 1 - f_{2} \right)
\,f_{3}\,\left( 1 - f_{4} \right) \,f_{5}
\end{eqnarray*}
Using (\ref{boltzmann-jacobian}) the Jacobian may be calculated\footnotemark[3]
\begin{eqnarray*}
J & = & {\hbox{circ}} {\Huge [}

{{\left( 1 - f \right) }^2}\,{f},- {{\left( 1 - f \right) }^2}\,{f}  ,
  {{\left( -1 + f \right) }^2}\,{f^2} ,
\\
& & \left( 1 - 2\,f \right) \,{{\left( 1 - f \right) }^2}\,{f},
  {{\left( 1 - f \right) }^2}\,{f}\,\left( -1 + 2\,f \right) , \\
& & -{{\left( 1 - f \right) }^2}\,{f^2}  {\Huge ]}

\end{eqnarray*}
Using (\ref{circulant-eigenvalues}), the eigenvalues of J may be directly
calculated
\begin{eqnarray*}
\lambda_0 & = & 0

\\
\lambda_1 & = & 0

\\
\lambda_2 & = & 2\,\epsilon \,\left( 1 + \epsilon  \right) \,{{\left( -1 + f
\right) }^3}\,{f}

\\
\lambda_3 & = & -6\,{{\left( 1 - f \right) }^2}\,{f^2}

\\
\lambda_4 & = & 2\,\epsilon \,\left( 1 + \epsilon  \right) \,{{\left( 1 - f
\right) }^3}\,{f}
\\
\lambda_5 & = & 0
\end{eqnarray*}
There are only three zero eigenvalues, so the deterministic FHP-type
lattice-gas model
possesses only three invariants: the total mass and the two components of
momentum.
The methodology of successively switching between left and right-handed
collision tables
is therefore justified, at least in the Boltzmann limit.  Switching between
left and
right-handed collision tables is done on the CAM-8 since there is no additional
time or
memory cost (per module) incurred in using multiple tables.  However, since
storing
multiple lookup tables costs additional memory on the CM-5, in practice we use
only a
single collision table, and therefore, our CM-5 simulations use a chiral
lattice-gas.  In fact,
the Von Karman street simulation in \S\ref{gallery} presented in
figure~\ref{vk_street} is an
example of a chiral lattice-gas.

\subsection{Single Particle Multispeed Fermi-Dirac Distribution Function}

It is essential to verify that in the macroscopic limit, the cellular automaton
equation of
motion (\ref{eq:cellular-update}) leads to Navier-Stokes hydrodynamics.   To
verify this, we
begin with the most general form of the single particle distribution function,
appropriate for
even multispeed lattice-gases: the Fermi-Dirac distribution.   Fundamentally,
this arises
because the individual digital bits used to represent particles  satisfy a
Pauli-exclusion
principle.  Therefore, the distribution must be written as a function  of the
sum of scalar
collison invariants, $\alpha + \beta e^\sigma_{ai} v_i + \gamma
\varepsilon_\sigma$,
implying the following form
\begin{equation}
\label{fermi-dirac-distribution}
f^\sigma_a = \frac{1}{1+e^{\alpha + \beta e^\sigma_{ai} v_i + \gamma
\varepsilon_\sigma}
}
\end{equation}
Using the identities in the appendix, an expansion to fourth order of
(\ref{fermi-dirac-distribution}) about zero velocity, close to that previously
calculated by
Chen {\it et al.} \cite{chen-jun91}, is the following
\begin{eqnarray*}
f^\sigma_a & = & d_\sigma \\
& & - d_\sigma (1-d_\sigma) \beta_1 e^\sigma_{ai} \frac{v_i}{c} \\
& & -\frac{1}{2} d_\sigma (1-d_\sigma) (\alpha_2 + \gamma_2

\varepsilon_\sigma)\frac{v^2}{c^2}\\
& & + \frac{1}{2} d_\sigma (1-d_\sigma)(1-2d_\sigma)\beta_1^2

e^\sigma_{ai}e^\sigma_{aj}\frac{v_i v_j}{c^2}\\
& & - \frac{1}{2}d_\sigma (1-d_\sigma) \beta_3 e^\sigma_{ai} \frac{v_i
v^2}{c^3} \\
& & + \frac{1}{2} d_\sigma (1-d_\sigma)(1-2d_\sigma)\beta_1 (\alpha_2 +
\gamma_2

\varepsilon_\sigma)e^\sigma_{ai}\frac{v_i v^2}{c^3}\\
& & - \frac{1}{6} d_\sigma (1-d_\sigma)(1-6d_\sigma + 6 d_\sigma^2)\beta_1^3

e^\sigma_{ai}e^\sigma_{aj}e^\sigma_{ak}\frac{v_i v_j v_k}{c^3}\\
& & + O(v^4)
\end{eqnarray*}
where $d_\sigma = f^\sigma_a \mid_{v=0}$.  The coarse-grain averaged dynamics
depend
on the following dynamical variables.

\\

Particle number density:
\begin{equation}
\label{density}
m\sum_{a,\sigma} f^\sigma_a = \rho,
\end{equation}
Momentum density:
\begin{equation}
\label{momentum}
mc\sum_{a,\sigma} e^\sigma_{ai} f^\sigma_a = \rho v_i,
\end{equation}
Moment density flux tensor:
\begin{equation}
\label{momentum-flux}
mc^2\sum_{a,\sigma} e^\sigma_{ai} e^\sigma_{aj} f^\sigma_a = \Pi_{ij}.
\end{equation}
Total energy density, half the trace of the momentum flux tensor:
\begin{equation}
n \varepsilon = \frac{1}{2}\hat{\Pi}_{ii}.
\end{equation}
Pressure tensor, $\hat{P}$:
\begin{equation}
\label{pressure-tensor}
\hat{P}_{ij} = m \sum_{a,\sigma} f^\sigma_a \left(c
e^\sigma_{ai}-v_i\right)\left(c

e^\sigma_{aj}-v_j\right).
\end{equation}
Heat flux, $\q$:
\begin{equation}
\label{heat-flux}
q_i = m \sum_{a,\sigma}  f^\sigma_a \left(c e^\sigma_{ai}-v_i\right)^2\left(c

e^\sigma_{aj}-v_j\right).
\end{equation}
In equilibrium, the cellular automaton dynamical equation
(\ref{lattice-boltzmann-equation}) reduces to

\begin{equation}
\label{equil-lattice-boltzmann-equation}
\partial_t f^\sigma_a + c e^\sigma_{ai}\partial_i f^\sigma_a = 0.
\end{equation}
(\ref{equil-lattice-boltzmann-equation}) implies 3 conservation equations.  To
obtain these
equations, the identities for isotropic lattice vectors given in the appendix
are necessary.
Using (\ref{density}) and (\ref{momentum}) in
(\ref{equil-lattice-boltzmann-equation}) gives
continuity (mass conservation):
\begin{equation}
\partial_t \rho + \partial_i (\rho v_i) = 0.
\end{equation}
Using  (\ref{momentum}) and  (\ref{momentum-flux}) in
(\ref{equil-lattice-boltzmann-equation}) gives the Navier-Stokes equation
(momentum
conservation):
\begin{equation}
\label{Navier-Stokes}
\partial_t (\rho v_i) +  \partial_j (\rho g v_i v_j)  = - \partial_i p + \eta
\partial^2 v_i.

\end{equation}
Using  (\ref{pressure-tensor}) and (\ref{heat-flux}) in
(\ref{equil-lattice-boltzmann-equation})
gives the heat equation (energy conservation):
\begin{equation}
\label{energy-equation}
\partial_t (n\varepsilon ) + \partial_i ( n\varepsilon v_i) +
\frac{1}{2}\partial_i q_i

	+ \partial_j (v_i P_{ij}) = 0,

\end{equation}
where
\begin{equation}
\label{pressure}
p = (\gamma-1) (n\varepsilon - \frac{1}{2} \rho g v^2)
\end{equation}
and $\gamma=\frac{C_p}{C_v}$ or $\gamma -1 = \frac{2}{D}$. Although the
lattice-gas
may in principle be comprised of an indefinite number of speeds,  from
(\ref{pressure}) we
see that the pressure depends upon the square of the bulk  velocity, {\it i.e.}
it is the
difference of the total internal energy of the lattice-gas minus the bulk
kinetic energy.  In a
single speed lattice-gas, this kind of velocity dependence is  anomalous and is
a well
known deficiency of the lattice-gas.  However, for a multispeed  lattice-gas
the existence
of this term takes on a physical interpretation.  For a classical  ideal gas,
the pressure is
proportional to both the sound speed squared and the  temperature
\begin{equation}
\label{ideal-gas-law}
p = \rho c_s^2 = n k_B T.
\end{equation}
Since $\rho = m n$, equating (\ref{pressure}) with the ideal gas law
(\ref{ideal-gas-law})
gives the total internal energy in terms of the bulk kinetic energy
$\frac{1}{2} m v^2$ and
the local particle thermal energy $k_B T$
\begin{equation}
\label{kinetic+heat_energy}
\varepsilon  = \frac{g}{2} m v^2 + \frac{k_B T}{\gamma-1} .
\end{equation}
The Navier-Stokes equation (\ref{Navier-Stokes}), the pressure
(\ref{pressure}), and the
total energy (\ref{kinetic+heat_energy}) all explicitly have a factor $g$ in
them\footnotemark[4]. If $g=1$, then a multispeed lattice-gas automata would
exactly
solve the ideal fluid equations where
the physical interpretation of the total internal energy would then be that it
partitions into a
bulk motion term, or kinetic energy, and a fluctuating motion term, or random
heat energy
associated with a certain gas temperature.  Therefore, the multispeed
lattice-gas
calculation, in the Boltzmann limit, would exactly agrees with classical
kinetic gas theory,
see the expression for the partial  pressure of an electron gas given by Li and
 Wu
\cite{Li-jun93}. The factor $g$ approachs one only as the number of speed in
the
lattice-gas model becomes large.  However, for a small number of speeds, a
rescaling of
the variables recovers the exact dynamics.
A similar observation has  been made by Teixeira \cite{teixeira-92} in his
investigation of
multispeed lattice-gas  automata models.

\subsection{Hexagonal Lattice}
\label{hexagonal-lattice}

In a hexagonal lattice there are six lattice vectors which we enumerate by
the following
convention
\begin{equation}
\label{moment-state-convection}
\ea = -\left(\sin\frac{\pi a}{3}, \cos\frac{\pi a}{3}\right),
\end{equation}
where $a=1,2,\dots,6$. The spatial coordinates of the lattice sites may be
expressed as
follows
\begin{equation}
\x_{ij} = \left( \frac{\sqrt{3}}{2} j , i - \frac{1}{2}(j \bmod 2) \right)
\end{equation}
where $i$ and $j$ are rectilinear indices which specify the data memory array
location
used to store the lattice-gas site data.  For a multispeed lattice-gas it is
necessary to shift
data more than one lattice length.  Let $s=(j \bmod 2)(r \bmod 2)$.  Given a
particle at site
$(i,j)$,  it may be shifted to a site  $r$ lattice units away to a remote site
$(i',j')$ by the
following mapping
\begin{eqnarray}
(i',j')_1 & = & \left( i+\frac{r+1}{2}-s ,  j - r \right) \\
(i',j')_2 & = & \left( i-\frac{r}{2}-s ,  j - r \right) \\
(i',j')_3 & = & \left( i-r ,  j  \right) \\
(i',j')_4 & = & \left( i+\frac{r+1}{2}-s ,  j + r \right) \\
(i',j')_5 & = & \left( i-\frac{r}{2}-s ,  j + r \right) \\
(i',j')_6 & = & \left( i+r ,  j  \right)

\end{eqnarray}
where $(i',j')_a$ denotes the shifted site, {\it i.e.} $(i,j) \rightarrow
(i',j')$ with a shift along
vector $\vec{\bf r} = r\e_a$ and where division by 2 is considered integer
division.

These streaming relations are useful for implementing a lattice gas in a
structured
language such as the C-language. Our implementation on the CM-5 in the
C-language and
DPEAC use these relations for all address computations. In these streaming
relations, the
modulus operator is base 2 because a two-dimensional hexagonal lattice embedded
into a
square three-dimensional mesh is pleated.

The simplest way to see this embedding is to define $z = (j \bmod 2)$.
Therefore the third
dimension along the z-axis is narrow, only one lattice distance wide.   Half of
the lattice
sites are at $z=0$ and the other half are at $z=1$.  This divides the
hexagonal lattice into
two sublattices that we refer to as {\it pleat 0} and {\it pleat 1}.
Table~\ref{stream-vectors}
lists the components of the data translation vectors, or kicks, for each
stream direction,
$a=1,2,\dots,6$, for both pleats.  This kick table was used for the CAMForth
implementation on the CAM-8 and the C* implementation on the CM-5. This is
equivalent
our general  streaming relations for the case when $r=1$.  The usefulness of
this kind of
embedding is that if the data for each one of the sublattices  is rendered for
display, it can
be drawn in simple raster form and fluid structures will  appear correctly,
{\it i.e.} a sound
pulse will appear circular.

\section{The Cellular Automata Machine CAM-8}
\label{the-cam-8}

The cellular automata machine CAM-8 architecture devised by Norman Margolus of
the
MIT Laboratory for Computer Science \cite{margolus-90b,margolus-93} is the
latest in a
line of cellular automata machines developed by the Information Mechanics Group
at MIT
\cite{toffoli-84b,toffoli-87,margolus-86}. It is optimized for performing
lattice-gas
simulations.  The CAM-8 architecture itself is a simple abstraction of lattice
gas dynamics.
Lattice gas data streaming and collisions are directly implemented in the
architecture.  The
communication network is a cartesian three-dimensional mesh.  Crystallographic
lattice
geometries can be directly embedded into the CAM-8.  Each site of the lattice
has a
certain number of bits (a multiple of 16) which we refer to as a ``cell''.
Each bit of the cell,
or equivalently each bit plane of the lattice, can be translated through the
lattice in any
arbitrary direction.  The translation vectors for the bit planes are termed
``kicks''.  The
specification of the x,y, and z components of the kicks for each bit plane (or
hyperplane)
exactly defines the lattice.  An interesting property of the architecture is
that the kicks can
be changed during the simulation. Therefore, the data movement in the CAM-8 can
be
quite general.  Once the kicks are specified, the coding of the lattice-gas
streaming is
completed. In effect, the kicks determine all the global permutations of the
data.

Local permutations of data occur within the cells.  These permutations are the
computational metaphor for physical collisions between
particles\footnotemark[5].  All
local permutations are implemented in look-up tables.   That is, all possible
physical
events with a certain input configuration and a certain output  configuration
are
precomputed and stored in SRAM, for fast table look-up.  The width of  the
CAM-8 look-up
tables are limited to 16-bits, or 64K entries.  This is a reasonable width
satisfying the
opposing considerations of model complexity versus memory size limitations  for
the
SRAM.  Site permutations of data wider than 16-bits must be implemented in
several
successive table look-up passes.  Since the look-up tables are double buffered,
a  scan of
the space can be performed while a new look-up table is loaded for the next
scan.

Figure~\ref{cam-8_system} is a schematic diagram of a CAM-8 system.  On the
left is a
single hardware module---the elementary ``chunk'' of the architecture.  On the
right is an
indefinitely extendable array of modules (drawn for convenience as
two-dimensional, the
array in normally three-dimensional).   A uniform spatial calculation is
divided up evenly
among these modules, with each module simulating a volume of up to millions of
fine-grained spatial sites in a sequential fashion.  In the diagram, the solid
lines between
modules indicate a local {\it mesh} interconnection.  These wires are used for
spatial data
movements.  There is also a tree network (not shown) connecting all modules to
the
front-end  host, a SPARC workstation with a custom SBus interface card, that
controls the
CAM-8.  It downloads a bit-mapped pattern as the initial condition for the
simulations. It
also sends a ``step-list'' to the CAM-8 to specify the sequence of kicks and
scans that
evolve the lattice-gas in time.  One can view the lattice-gas simulation in
real-time since a
custom video module captures site data for display on a VGA monitor, a useful
feature for
lattice-gas algorithm development, test and evaluation.  The CAM-8 has built-in
25-bit
event counters so that measurements can be done in real-time without slowing
the
lattice-gas evolution.  We have used this feature to do real-time coarse-grain
block
averaging of the lattice-gas number variables and to compute the components of
the
momentum vectors for each block.  The amount of coarse-grained data is
sufficiently small
that it can be transferred back to the front-end host for graphical display as
an evolving
flow field within an X-window.  See
figures~\ref{convection_rolls},~\ref{kh_fhp2}
and~\ref{vk_street} for example flow fields.

\section{The Connection Machine CM-5}
\label{cm-5}

	The Thinking Machines Corporations's CM-5 is a massively parallel computer
that
contains up to 16384 processing nodes\cite{TMC-nov92}\footnotemark[6].
Figure~\ref{node} shows a  processing node consisting of a SPARC CPU, 32 Mbytes
of
memory and 4 Vector processing units. These processing nodes are all connected
via a
``fat-tree" communications net that allows fast inter-node communication. These
processing nodes are controlled by a front-end host computer which is a
modified SUN
workstation. The SPARC processor on each node issues instructions to the vector
units
and performs most bookkeeping tasks while the vector units perform arithmetic
and logical
operations on the data.  Each vector unit has a peak rate of 32 million 64-bit
ops  (floating
point or integer) for a combined total of 128 Mops/node. Each node's memory is
divided
into 8 Mbyte banks, one for each vector unit. Each vector unit has it's own
independent
128 Mbyte/sec path to memory for a combined memory bandwidth of 512 Mbyte/sec
for
each node.  The CM-5 at the Army High Performance Computing Research Center in
Minneapolis, Minnesota contains 544 nodes for a total of 16 Gb of memory and 64
Gops of
peak processing speed.

We have also implemented a lattice gas simulator on the CM-5 in a multiple
instruction
multiple data (MIMD) style. The CMMD message passing library is used for
inter-node
communication and host-node interaction. In order to get the highest possible
performance
we  explicitly manipulate the vector units on each node using their assembler
language
known as DPEAC. To ease the burden of hand coding the vector units a macro
package
known as GCC/DPEAC is used. This package uses features available in the GNU C
compiler to issue assembler language instructions from ANSI C and simplifies
matters
considerably.

We partition the problem space into equally sized rectangular units.  Each
processing
node is responsible for updating one of these rectangular units. This
partitioning allows one
to send a small number of long messages to connect the space together.
Inter-node
communication is only necessary along one of the axes of the problem space.
Since the
inter-node communications network is optimized for long message lengths we
expect that
this partitioning will make the effective use of available communications
bandwidth. Within
a processing node, each of the 4 vector units is responsible for updating it's
quarter of the
space. Communication between each vector unit's 8 Mbyte bank of memory is
mediated
by the SPARC processor.

There are two distinct phases of a lattice gas update cycle. The first phase is
the collision
phase where particles can interact. The second phase involves streaming of the
bits to
their new locations, consistent with their velocity and the lattice on which
the simulation is
being performed. In most lattice gas models all collisions can happen
concurrently and all
sites can stream their data concurrently as well.  The collision phase can be
handled via
look up tables (LUT's) for 16 bit sites. The LUT is attractive in that it can
be an extremely
simple and fast update mechanism.

We have distributed the LUTs throughout the machine, indeed each vector unit
has it's
own copy of the LUT. Figure \ref{mem_layout} shows the memory layout on each
node.
During the collision phase each vector unit fetches all the sites in it's
partition of the
problem space and runs them through it's copy of the LUT. Since each vector
unit has it's
own independent 128 Mbyte/sec data path to a bank of memory, this operation can
be
performed extremely rapidly. With this high degree of parallelism the LUT
operation
consumes a small fraction of the time necesary to update the space. As the
number of bits
of site data grows beyond 16 [64K entries] the LUT's begin to consume too much
memory.
For models that involve larger quantities of site data ({\it i.e.} \# bits $>$
20) other methods
involving LUT compression/decompression need to be used for the collision
phase.\cite{henon-92}

The streaming phase is more complex. The approach taken here is to hold the
address of
each bit's destination in a pre-computed table. These tables may be computed
in ordinary
C, which is advantageous for changing from one model to another.

Additionally, potentially complex addressing calculations are performed only
once, during
initialization. Before a site can be updated, a communication phase must take
place so
that each site can access all it's neighbors. Communication must take place
across node
and vector unit boundaries. The communication is done so that every site has
access to
all it's neighbor values on one vector unit's 8 Mbyte bank of memory. To update
a
particular site each vector unit first loads all the neighbors of the site to
be updated from
the appropriate areas. Then the relevant bits from each neighbor site are
extracted
sequentially to build up the new site value. The new site value is then written
to memory.

After implementing a lattice-gas simulator with the above considerations in
mind we find
that we can achieve update rates on the order of 550 Msites/sec on a 256 node
partition of
the CM-5 for the FHP gas model. This timing was done
using a 32K $\times$ 2K lattice. We find that the longer the system is across
each node the
greater the performance realized. This is due to the fact that long
system sizes across each node increase the fraction of sites in the interior
of each vector unit that do not need to communicate with sites on adjacent
vector units or processing nodes.

\section{Gallery of Computational Results}
\label{gallery}

\subsection{Rayleigh-B\'enard Convection on the CAM-8}

A well known fluid instability of a thermohydrodynamic system is
Rayleigh-B\'enard
convection\cite{normand-jul77,cohen-jan84}.  Rayleigh-B\'enard convection is
popular
because one can observe the onset of order and then the transition to chaos in
the flow
patterns \cite{careri-84,baker-90}.

Our implementation of the  two-speed hexgonal lattice-gas with a rest particle,
includes
gravitational forcing, free-slip and no-slip boundaries which may be oriented
horizontally,
vertically, or inclined $\pm60^\circ$, and heating and cooling sites in order
to model
temperature controlled boundary surfaces. This has been encoded within the site
data
space of 16-bits per site for simple implentation on the CAM-8.  The ability of
encoding
such complex dynamics within 16-bits is one of the remarkable aspects of the
lattice-gas
formalism in terms of efficient memory use affording us the ability to do flash
updating from
prestored collision tables.  $98\%$ of the $2^{16}$ entries in the collision
tables are used
({\it i.e.}, not the identity) in this model.  Similar lattice gas models have
been implemented
by Burges and Zaleski \cite{burges-87} and by Chen {\it et al.}
\cite{chen-jun91} and Ernst
and Das\cite{ernst-jsp66-92}.

To optimize the collision frequency between the fast and slow particles we have
chosen
their momenta to be of unit value.  That is, the slow particles have unit mass,
$m_1 = 1$
and the fast particles have half the mass, $m_2 = 1/2$. In this way, $p_1 = p_2
= 1$ and
their energies are $E_1 = 1$ and $E_2 = 2$. With this convention we have the
usual
FHP-type collisions \cite{frisch-86} between the different speed particles
while conserving
mass, momentum, and energy.  These include head-on 2-body collisions,
three-body
collisions, collisions with spectators, etc.  Grosfils, Boon and Lallemand have
introduced a
three-speed thermohydrodynamic gas with speeds $1,\sqrt{3},2$ and a rest
particle
\cite{grosfils-jsp-93}. With this 19-bit model, efficient collisional mixing
can occur with all
particle having the same unit mass.  Since the particles may now carry
different units of
energy, in addition to the equation of continuity and Euler's equation, in this
system we
have an energy transport equation.

\subsection{Kelvin-Helmholtz Instability on the CAM-8}

Another well known fluid instability is the Kelvin-Helmholtz shear instability.
Figure~\ref{kh_fhp2} shows the a simulation of a shear instability on a
hexagonal lattice
$4096\times2048$ in size with toroidal boundary conditions.  A momentum map is
overlayed on a vorticity map.  Clockwise vorticity is shaded red and
counter-clockwise
vorticity is shaded blue. The initial conditions for the simulation are very
uniform, see
Figure~\ref{kh_fhp2}a.  A gas density is chosen, in this case approximately 1/7
filling, and
two horizontal regions are set with uniform, but opposing flow directions.
That is, the
majority of the fluid, the background region, is set with a uniform flow
velocity of
approximately 0.4 $c_s$ (Mach 0.4) flowing to the right.  A narrow stripe 256
sites wide is
set in the center of the space flowing to the left at -0.4 $c_s$.  No
sinusoidal perturbation is
given to the counter-flow narrow stripe region as in previous lattice-gas
simulations\cite{Shimomura-90}.  No external forcing is applied during the
simulation run.
The only perturbation is caused by minor fluctations produced by the random
number
generator when producing a uniform fluid density.  After approximately 10,000
time steps,
the narrow horizontal center stripe forms a sinusoidal pattern.  The sinsoid's
amplitude
grows, form a wave that eventually breaks into several counter-rotating
vortices and the
two flow regions begin to substantially mix.  Figure~\ref{kh_fhp2} shows the
initial state of
the fluid and then the resulting states at 10,000 and 30,000 time steps.  By
$t=30,000$ the
formation of a wave is apparent. Eventually after 400,000 time steps, the fluid
attains a
uniform flow to the right after the system has equilibrated, exactly conserving
the
momentum in the initial configuration.

\subsection{Von Karman Streets on the CM-5}

Figure~\ref{vk_street} shows a simulation of vortex shedding from a flat plate
after 32,000
time steps on a hexagonal lattice $4096\times2048$ in size.  A momentum map is
overlayed on a vorticity map.  Clockwise vorticity is shaded red and
counter-clockwise
vorticity is shaded blue. A flat plate obstacle is placed in a channel of fluid
with a flow
directed towards the right of the figure. The fluid flow is forced by
completely
reconstructing the fluid's velocity distribution at each time step in a
``forcing strip'' at the
left of the channel. This forcing method prevents sound waves and other
disturbances from
propagating around the torus in the flow direction and overwhelming the flow
behavior.
The boundary conditions are effectively cylindrical.

The flow is started from a random distribution of particles at the appropriate
density with a net velocity close to that of the steady state flow. Since this
is not a true equilibrium starting condition, some transient behavior appears
in the form a
sound pulse that propagates down the channel. This pulse
is absorbed by the forcing strip. After 2000 time steps the system is
equilibrated with no
transient phenomena visible.  This equilibration time is very short compared
with the time
necessary for vortex development.
The cylindrical boundary condition appears to work extremely well, allowing one
to run the
simulation to long times.

\section{Discussion}

We have presented a theoretical description of lattice-gas automata in a
discrete
D-dimensional space.  Attention was focused on two-dimensional fluids for
numerical
simulation. We have implemented some lattice-gas fluids on a new prototype
cellular
automata machine, the CAM-8.  Identical models on the CM-5 were also
implemented.  To
illustrate lattice-gas dynamics in the macroscopic limit, several fluid
instabilities were
tested: (1) the Rayleigh-B\'enard convective instability just after the
critical Rayleigh
number has been reached in the system by suitable gravitational forcing and
temperature
gradient; (2) the Kelvin-Helmhotz shear instability; and (3) the Von Karman
vortex
shedding instability.

The strength of the CAM-8 in these simulations is that it is optimized
for fine-grained spatial calculations.  It can handle many
lookup-tables because of its double buffering.  It can perform data
streaming by spatial data shifts without slowing down the simulation.
For a multispeed system or a lattice-gas with long-range interactions,
large shifts are necessary.  The interaction neighborhood on the CAM-8
need not be local: data can be shifted to the nearest neighbor or a
distance thousands of sites away with the same computational effort.

The strengths of the CM-5 in these simulations are its software, size,
and flexibility.  CM-5 applications can be coded largely in standard
programming languages such as C and FORTRAN-90.  CM-5 machines offer
the possibility of simulating enormous problem spaces due to their
much larger memories.  In our implementation, streaming is done in the
CM-5 by address maps preloaded into memory.  Although this uses a lot
of memory, the flexiblity of precomputed addressing tables simplifies
the implementation of complex lattice geometries.

The main theoretical and computational points of this paper are:

1. Deterministic microscopic lattice-gas dynamics produce the correct
Navier-Stokes
hydrodynamics in the macroscopic limit where no randomness is used in the local
rules.
Treating individual digital bits as multispeed fermionic particles allows one
to simulate fluid
systems where mass, momentum, and energy are exactly conserved and where the
dynamics has a time-reversal invariance.

2. A workstation-scale CAM-8 prototype is an efficient, cost-effective
platform for lattice gas problems that rivals the capabilities of
extant parallel supercomputers.

3. The lattice gas simulation method may be directly ported to a
variety of parallel computer architectures: the facilities provided by
most parallel supercomputers are suitable for efficiently running
lattice gas applications.  Communication costs for running lattice-gas
automata simulations decrease as the problem being run increases in
size and can essentially be neglected for large systems.

In closing, it is interesting that the massively parallel Connection Machine 2,
to
date has achieved some of the best update rates for lattice-gases, for
example, 700-750 million sites per second update
rate for the FHP lattice-gas\cite{schen-90,boghosian-91}.  Building a multiple
instruction multiple data Connection Machine 5, although more focused
on general messaging, has improved upon its predecessor's performance
(for the FHP lattice-gas, currently the CM-5 could exceed 2 billion
updates per second on a 1024-node partition).  Although machines like
the CM-5 can solve a wider class of computational problems, we believe
that effort spent on exploring locally interacting automata models on
fine-grained architectures will lead to new practical methods for
accurately modeling physics.

\section{Acknowledgements}

JY would like to thank Dr. Bruce Boghosian for several discussions,
particularly those
concerning latttice-gas invariants.  Thanks expressed also to Donald Grantham
and Dr.
Robert McClatchey of Phillips Laboratory's Atmospheric Sciences Division for
their
support of the lattice-gas basic research initiative at our laboratory.
Finally, thanks go to
Dr. Marc Jacobs,  Air Force Office of Scientific Research, who promptly
arranged for
supercomputer time on the Connection Machine that has proved very valuable to
our
progress.

Research time on the Connection Machine has been supported by, or in part by
the Army
Research Office, contract number DAAL03-89-C-0038 with the University of
Minnesota
Army High Performance Computing Research Center at Minneapolis, Minnesota. Our
CM-5 code for lattice gas simulation will be deposited there for use by
interested parties.

\newpage

\appendix{Identities for Isotropic Lattice Tensors:
$\prod_{k=1}^{2n}\e^\sigma_k$}

Our momentum states are $\e^\sigma_a$, where $a=1,2,\dots,B_\sigma$, and where
$\sigma$ denotes the particle speeds. The momentum state-space per particle
speed has
cardinality $B_\sigma$.
Lattice summations over odd powers of $\e$ must vanish by symmetry.  The
following
identities, listed up to the fourth moment, hold for arbitrary values of
$B_\sigma$ and
spatial dimension $D$ \cite{wolfram-86}
\begin{equation}
\sum_a e^\sigma_{ai}  =  0\end{equation}
\begin{equation}
\sum_a e^\sigma_{ai}e^\sigma_{aj}  =  \frac{B_\sigma}{D}
\delta_{ij}\end{equation}
\begin{equation}
\sum_a e^\sigma_{ai}e^\sigma_{aj}e^\sigma_{ak} =  0
\end{equation}
\begin{equation}
\sum_a e^\sigma_{ai}e^\sigma_{aj}e^\sigma_{ak}e^\sigma_{al}  =
\frac{B_\sigma}{D(D+2)}(\delta_{ij}\delta_{kl}+\delta_{ik}\delta_{jl}+\delta_{il}\delta_{jk}) .
\end{equation}

\appendix{C* Implementation}
\label{c*}

A parallel version of the C-language developed by Thinking Machines Corporation
is the
C* language\cite{TMC-nov90}.  This is a well developed language in spirit very
close to its
predecessor -- it is as concise as the C-language itself.  It offers many
useful constructs
making the coding of algorithms for parallel data very efficient.   As typical
of most parallel
languages, an array operation is handled in a single instruction --- for the
most part
programming loops do not appear in the code.  Most parallel computation is
achieved by
data movement.  The geometry of the problem is specified at the onset by
defining your
the data structure's {\it shape}.  This is usually a D-dimensional array with a
certain size in
each dimension.  The shape definition defines all the needed communication
topology for
the compiler.  It is possible to declare boolean shapes in C*.  We have used
this feature to
encode each bit plane of the lattice-gas.  This is a convenient feature of the
language
making efficient use of memory.  Normally, in a lattice-gas code, one must
extract and
insert individual bits at the lattice sites.  The option of working directly
with boolean arrays
has therefore simplified the coding effort substantially.  If individual
elements of a parallel
array must be accessed, C* uses the syntax of parallel left indexing.  Right
indexing of
arrays is reserved for its usual C-language meaning.  We use right indexed
arrays to
represent the individual bit planes of the lattice gas.

We have implemented a two-dimensional hexagonal lattice embedded into a
three-dimensional mesh. This implementation is equivalent to our implemention
in
CAMForth on the CAM-8 and GCC/DPEAC on the CM-5.  Streaming of pleat 0 and
pleat 1
are coded separately.  We give a C* code fragment for this embedding.  Note
that the
comments to the right of the lines correspond exactly to the kick components
listed in
Table~\ref{stream-vectors}.  With a few C* lines of code one can completely
implement
hexagonal lattice-gas  streaming.  We have used the C* command {\it
to-torus-dim(destination pointer, source,  axis, distance)} to shift the bit
planes with toroidal
boundary conditions.  This is an efficient  communication routine for sending
data in a
regular fashion using grid communications.   The partitioning of the space
between
processors is handled completely by the C*  compiler.  We will see how
efficiently the
compiler does this partitioning in the discussion  to follow.

We have tested our C* implementation for different situations.  Given a certain
lattice size,
for example $1024\times2048$, with have found the performance of the CM-5 to
vary
linearly with the number of processing nodes.  This linear variation is
expected so long as
the lattice size is sufficiently large.  To determine a reasonable lattice
size, we have
performed repeated simulations with different lattice sizes but with a fixed
number of
processors.  The results obtained for a fixed 256-node partition of the CM-5 is
given in
figure~\ref{cs-embedded-hex}, in which we plotted simulation site update rates
for lattice
sizes $64\times 128, 128\times 256, \dots, 8192\times16384$.  For small lattice
sizes, the
performance is very poor, on the order of a million site updates per seconds.
This is
because the streaming is limited by processor to processor communication
bandwidth.  As
the lattice size increases, the number of sites interior to the node grows and
the number of
sites on the partition boundary deceases.  Consequently, the site update rate
continuously
improves with larger lattices.  The update rate asymtotically approaches about
25 million
site updates per second.  This is equivalent to approximately 100,000 sites
updates per
processing node.  This is roughly the maximum update rate achievable on a
SPARCstation
1.  A full 512-node partition has a peak rate of about 50 million site updates
per second,
which is about one quarter the speed of the CAM-8.

\newpage

\begin{table}
\caption{Streaming for 2D Hex Lattice Embedded a into 3D Mesh}
\label{stream-vectors}
\begin{tabular}{lccc}
{\bf Direction} & {\bf x} & {\bf y} & {\bf z}\\
\hline
{\sc Pleat 0}   &    &    &  \\

1	         &   0 & 0 &  1     \\
2		       &   0 &   -1 &  1     \\
3		       &   0 &   -1 &  0     \\
4		       &   1 &   -1 &  1     \\
5		       &   1 &   0 &  1     \\
6		       &   0 &   1 &  0     \\ \hline
{\sc Pleat 1}   &      &      &  \\

1		       &   -1 &   1 &  -1     \\
2		       &   -1 &   0 &  -1     \\
3		       &   0 &   -1 &  0     \\
4		       &   0 &   0 &  -1     \\
5		       &   0 &   1 &  -1     \\
6		       &   0 &   1 &  0

\end{tabular}
\end{table}

\newpage

\footnotetext[1]{We have written lattice-Boltzmann code in the parallel C-Star
language on
the Connection Machine-5.}

\footnotetext[2]{Lattice Boltzmann simulation  for three dimensional flows with
Reynolds
numbers of about 50,000 were presented in  June 1993  at the International
Conference
on Pattern Formation and Lattice-Gas  Automata sponsored by the Fields
Institute,
Waterloo, Canada.}

\footnotetext[3]{Actually, the  most straight foward way to calculate the
matrix elements of
J is to write a Mathematica  code.}

\footnotetext[4]{In the Boltzmann limit, the factor $g$ depends on the particle
speeds,
$c_\sigma$, and on the density distribution per speed, $d_\sigma$, by the
following
complicated expression
\begin{equation}
g = \frac{\sum_\sigma d_\sigma \sum_\sigma B_\sigma (\frac{c_\sigma}{c})^4
d_\sigma
(1-d_\sigma)(1-2d_\sigma)}
{\left[ \sum_\sigma B_\sigma (\frac{c_\sigma}{c})^2 d_\sigma (1-d_\sigma)
\right]^2}.
\end{equation}
}

\footnotetext[5]{Locally, the  CAM-8 is not limited to performing only
permutations, it can
do general mappings.   However, since we are interested in only particle
conserving
reversible dynamics,  permutations are sufficient.}

\footnotetext[6]{Currently  the largest CM-5 resides at Los Alamos National
Laboratory with
1024 Nodes.}

\newpage
\onecolumn

\epsffile{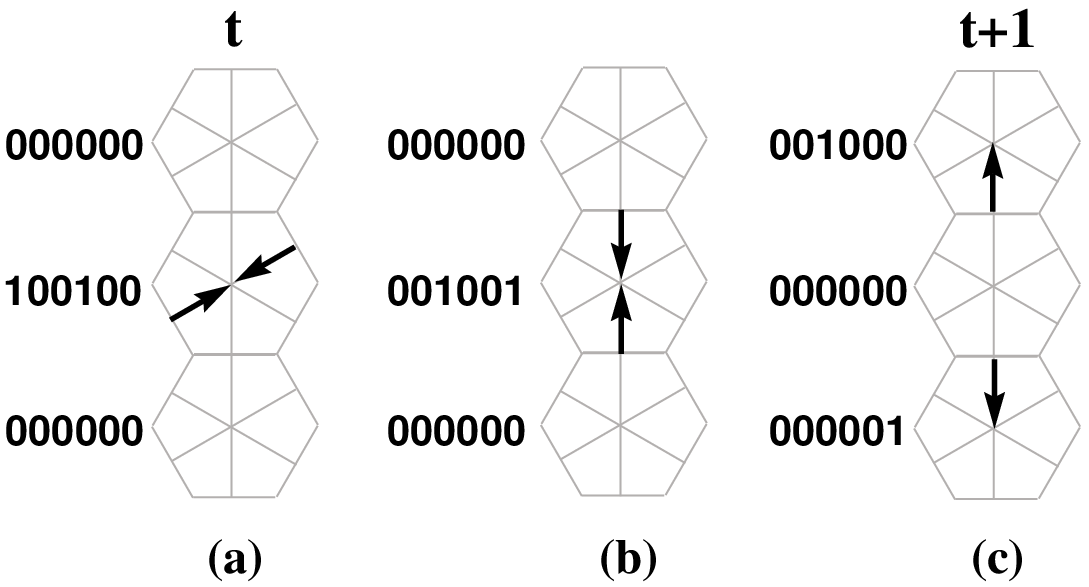}
\figure{Illustration on a hexagonal lattice of the two-step collision and
streaming process
required to complete a single time step: (a) initial configuration, (b)
collision by
permutation, and (c) streaming of particles.\label{collision_streaming}}

\epsffile{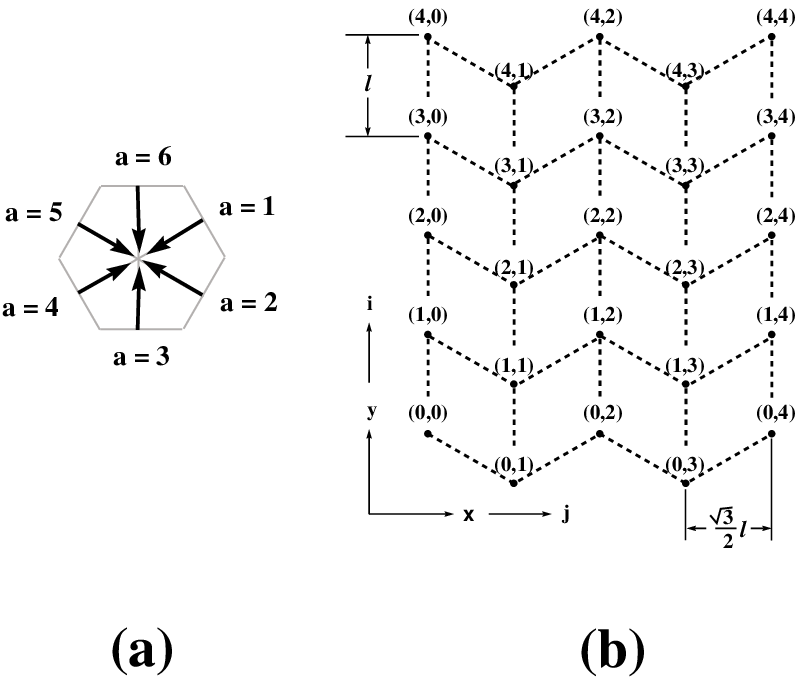}
\figure{(a) Lattice vector label convention;  (b) Hexagonal lattice convention
with lattice
directions $a=3$ up and $a=6$ down. Coordinates above the lattice nodes are
$(i,j)$
memory array indices.\label{hex_lattice}}

\epsffile{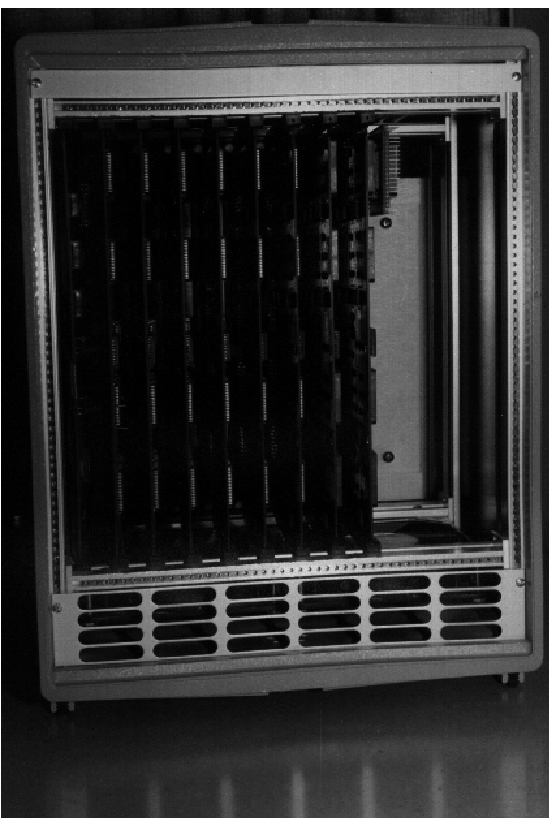}
\figure{MIT Laboratory for Computer Science cellular automata machine CAM-8.
This 8
module prototype can evolve a D-dimensional cellular space with 32 million
sites where
each site has 16 bits of data with a site update rate of 200 million per
second.
\label{cam-8_prototype}}

\epsffile{./figures/cam-8_system.eps}
\figure{ (a) A single processing node, with DRAM site data flowing through an
SRAM
lookup table and back into DRAM. (b) Spatial array of CAM-8 nodes, with
nearest-neighbor (mesh) interconnect (1 wire/bit-slice in each direction).
\label{cam-8_system}}

\epsffile{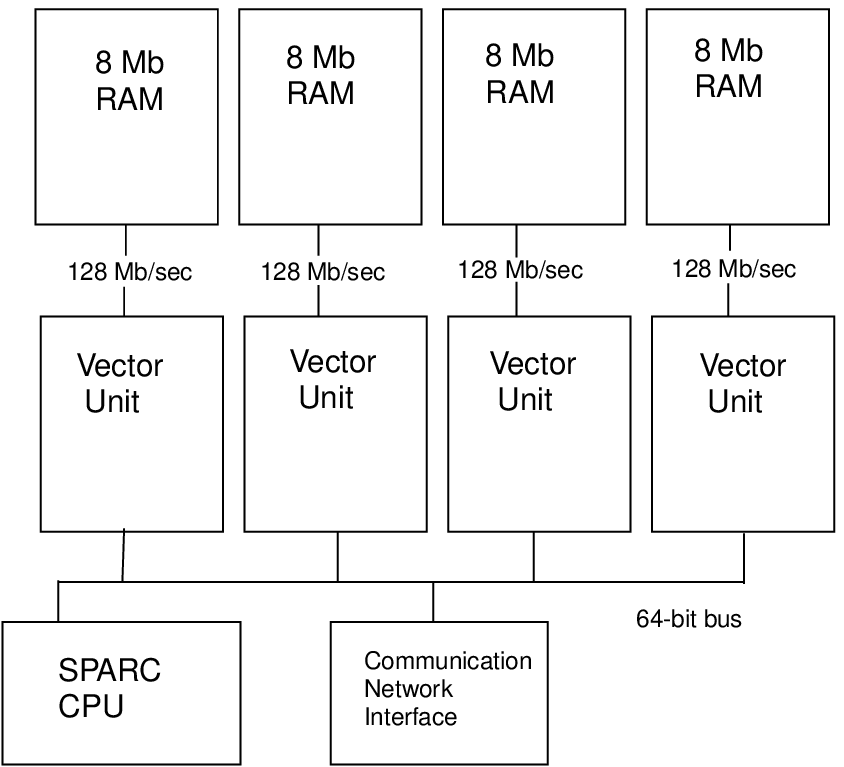}
\figure{CM-5 Node: SPARCCPU, 32 Mbytes of memory and 4 Vector processing
units.\label{node}}

\epsffile{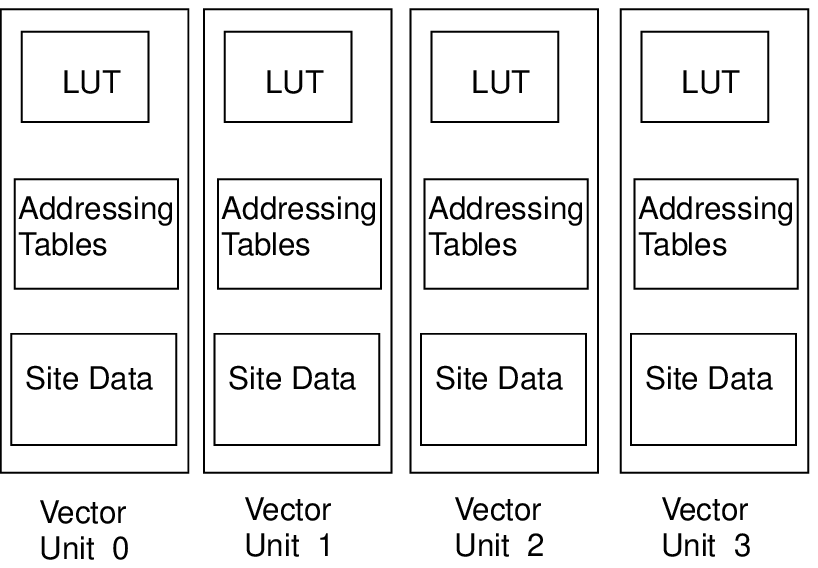}
\figure{Node Memory Layout\label{mem_layout}}

\epsffile{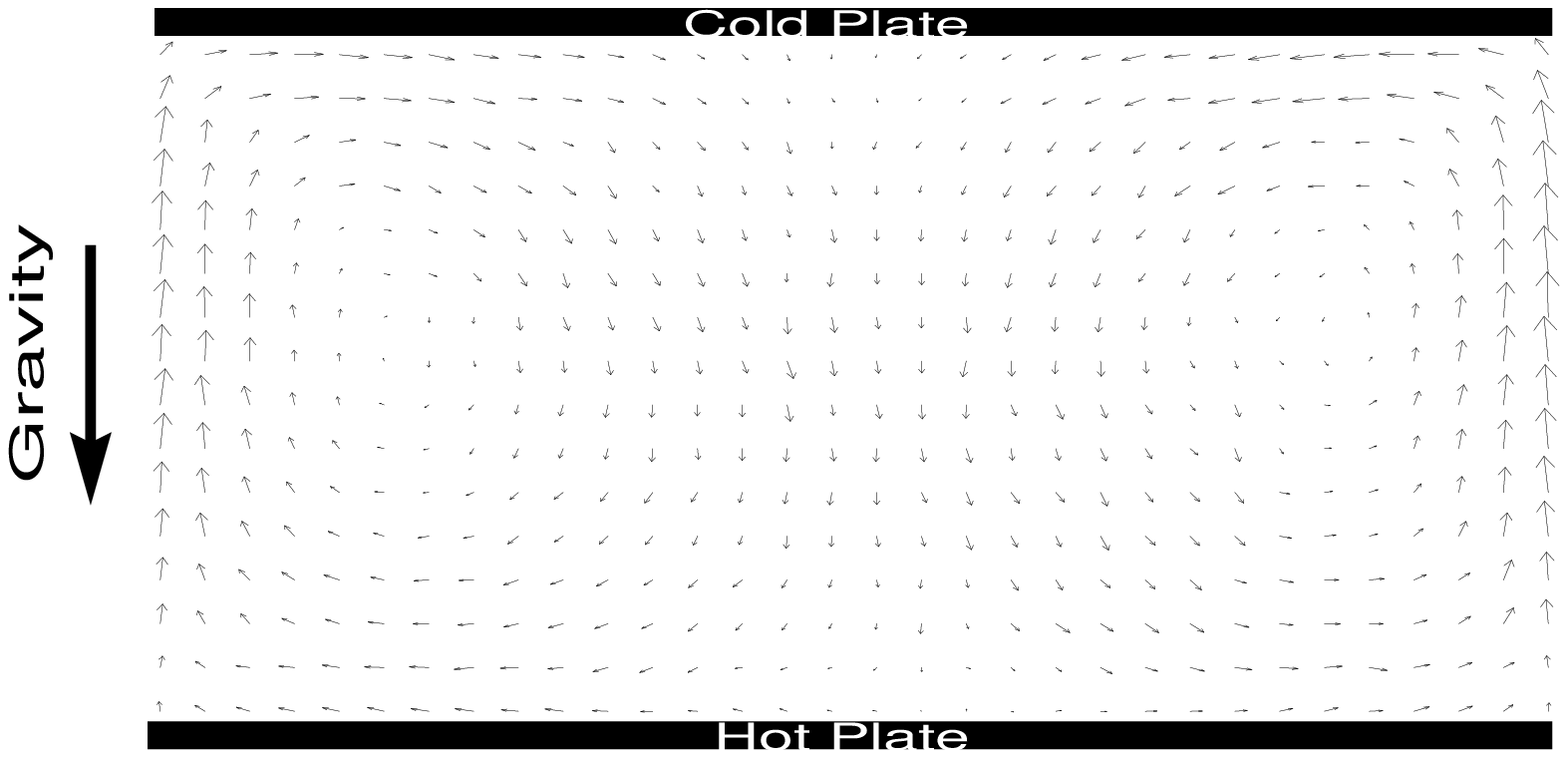}
\figure{Thermo 13-bit CAM-8 Experiment: Rayleigh-B\'enard convection cells at
the critical
Rayleigh number. Lattice Size: $2048\times1024$. Time Average: 100. Spatial
Average:
$64\times64$. Mass Density Fraction=1/5. Data presented at 50,000 time
steps.\label{convection_rolls}}

\epsffile{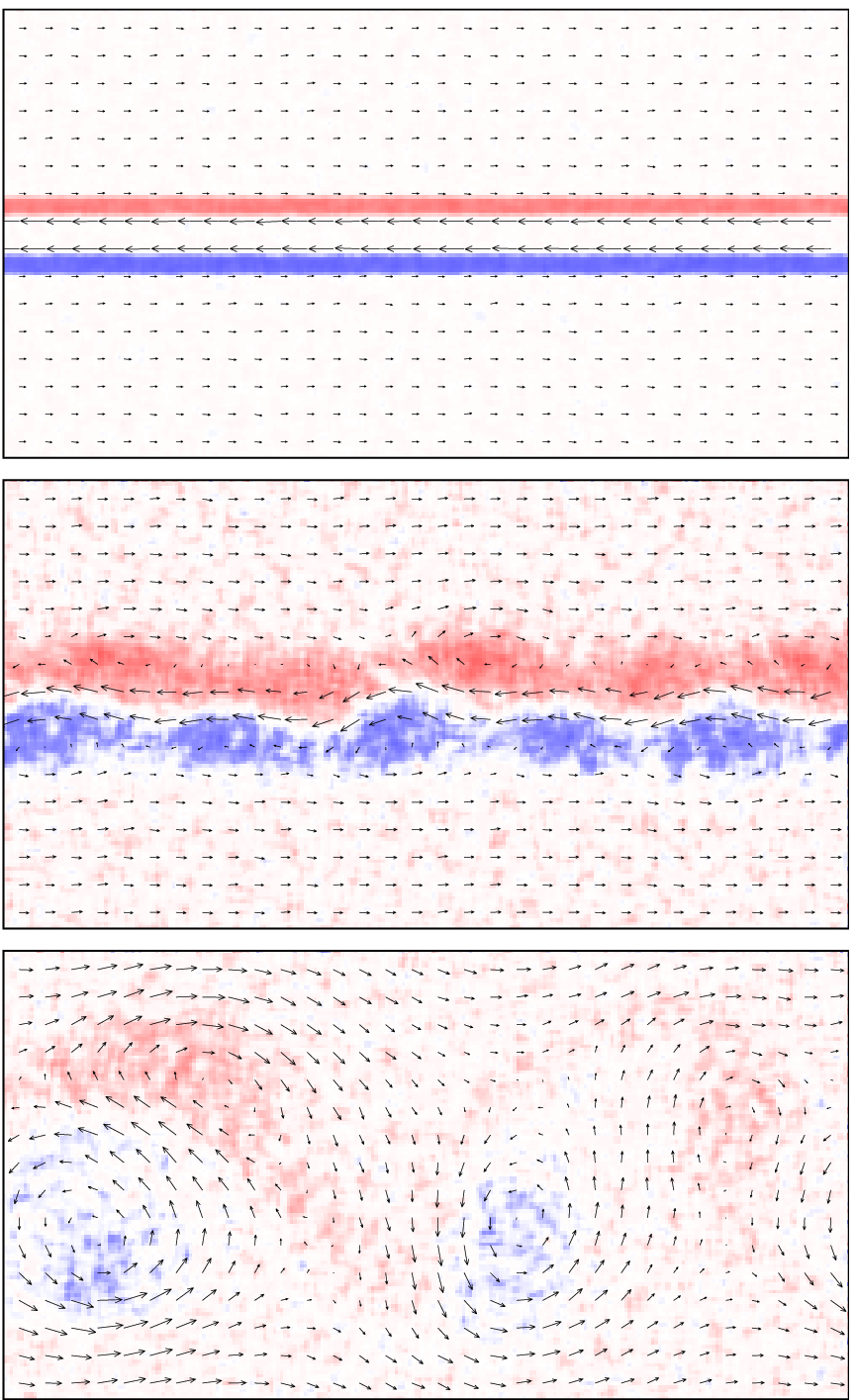}
\figure{Momentum and vorticity map of two-dimensional shear instability on the
CAM-8.
Lattice size of  $4096\times 2048$ with toroidal boundary conditions. Spacetime
averaging
over 128x128  blocks for 50 time steps.  FHP collisions with spectators and a
rest particle.
Data presented  at time steps 0, 10000, and 30000 with Galilean velocity
shift.\label{kh_fhp2}}

\epsffile{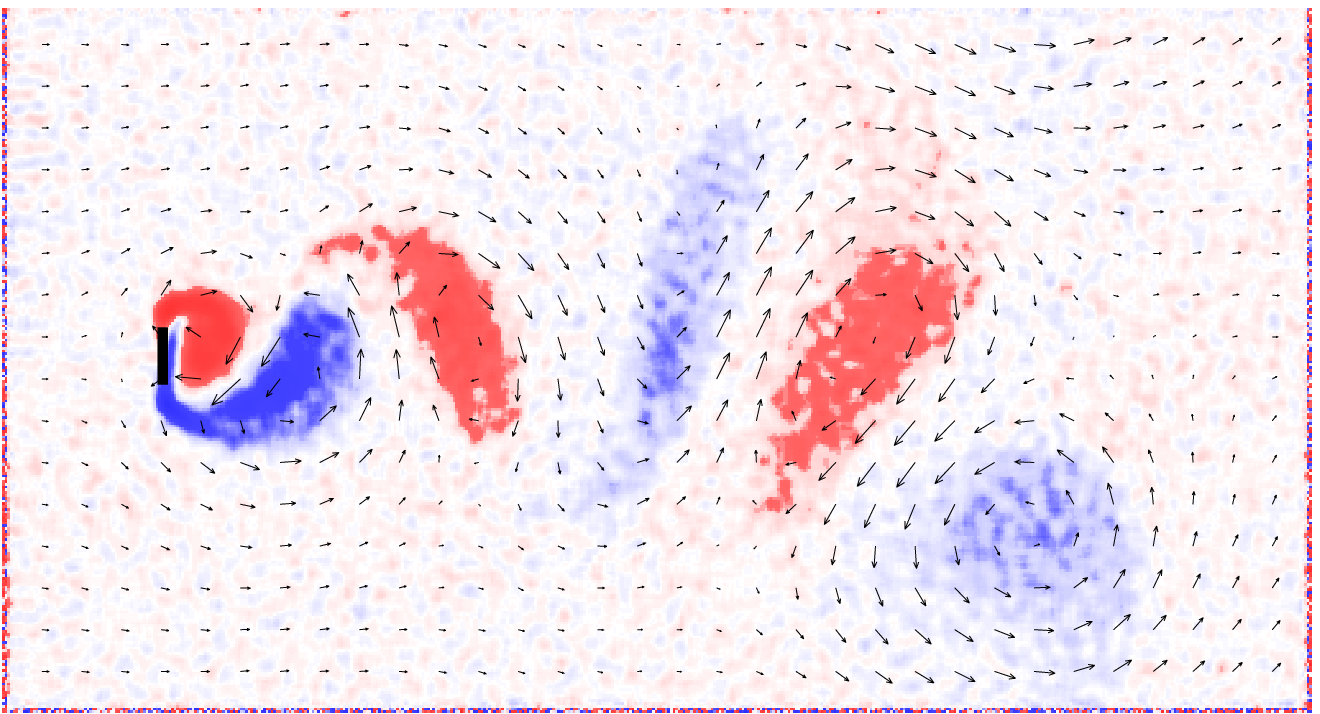}
\figure{FHP-II CM-5 Experiment: Von Karman Streets  Lattice Size:
$4096\times2048$.
Time Average: None. Spatial Average: 128x128. Mass Density Fraction=1/7. Data
presented at 32,000 time steps.\label{vk_street}}

\epsffile{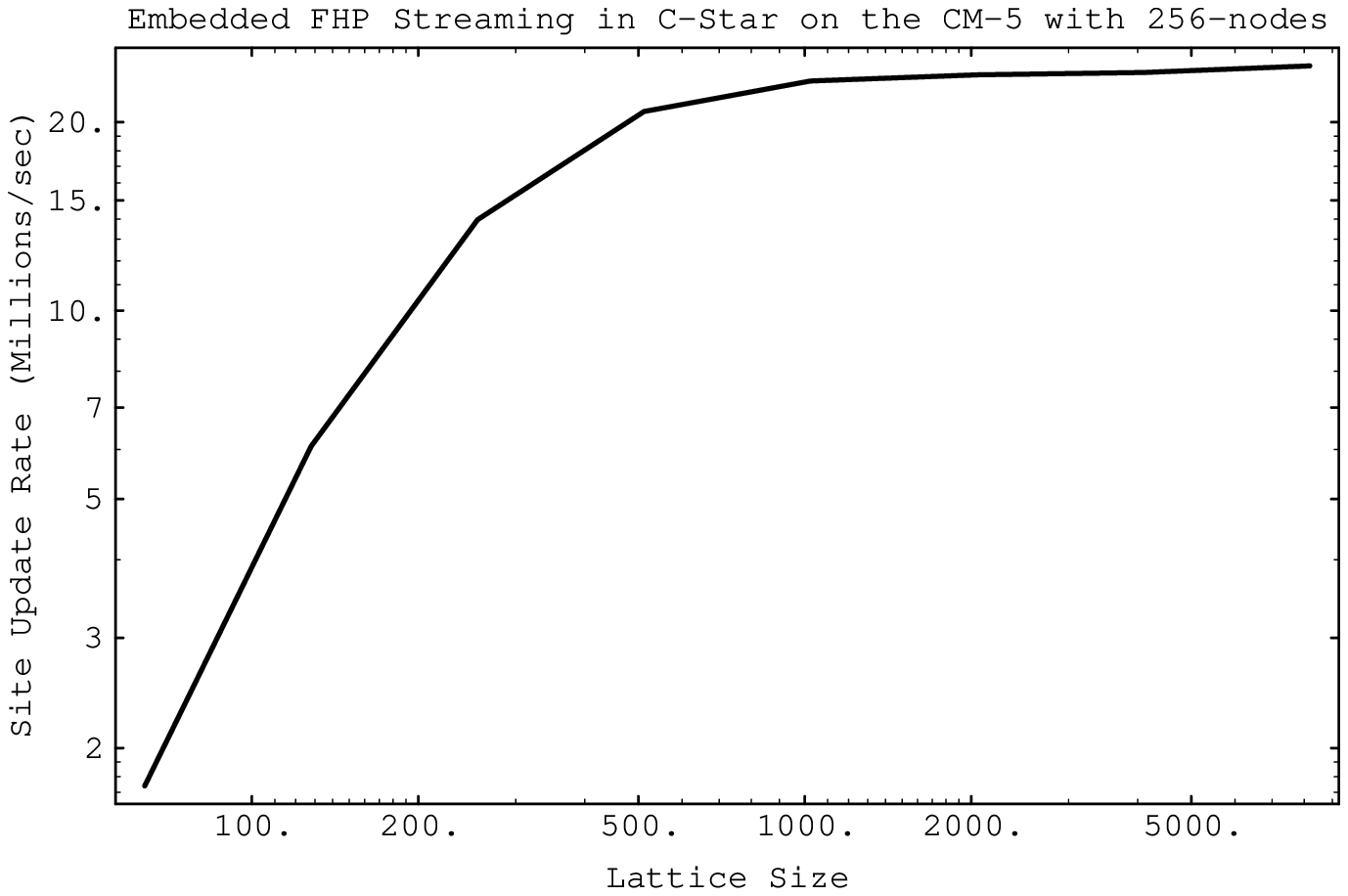}
\figure{Performance runs on a 256-node CM-5 for an FHP hexagonal lattice
embedded
into a 3D mesh. Performance significantly suffers by communication overhead for
small
lattice sizes.\label{cs-embedded-hex}}

\newpage
\twocolumn


\begin{thebibliography}{10}

\bibitem{rahman-oct64}
A.~Rahman.
\newblock Correlations in the motion of atoms in liquid argon.
\newblock {\em Physical Review}, 136(2A):405--410, 1964.

\bibitem{wolfram-83}
Stephen Wolfram.
\newblock Statistical mechanics of cellular automata.
\newblock {\em Reviews of Modern Physics}, 55(3):601--644, 1983.

\bibitem{wolfram-84}
Stephen Wolfram.
\newblock Universality and complexity in cellular automata.
\newblock {\em Physica}, 10D:1--35, 1984.

\bibitem{toffoli-87}
Tommaso Toffoli and Norman Margolus.
\newblock {\em Cellular Automata Machines}.
\newblock MIT Press Series in Scientific Computation. The MIT Press, 1987.

\bibitem{toffoli-84a}
Tommaso Toffoli.
\newblock Cellular automata as an alternative to (rather than an approximation
  of) differential equations in modeling physics.
\newblock {\em Physica}, 10D:117--127, 1984.

\bibitem{liboff-79}
Richard~L. Liboff.
\newblock {\em Introduction to the Theory of Kinetic Equations}.
\newblock Robert E. Krieger Publishing Company, 1979.

\bibitem{rivet-86}
J.P. Rivet and U.~Frisch.
\newblock Lattice gas automata in the boltzmann approximation.
\newblock {\em Comptes Rendus}, 302(II):p. 267, 1986.
\newblock In French. Translation appears in Lattice Gas Methods for Partial
  Differential Equations, SFI SISOC, Eds. Doolen et al., Addison-Wesley
  Publishing Co., 1990.

\bibitem{mcnamara-88}
Guy~R. McNamara and Gianluigi Zanetti.
\newblock Use of the boltzmann equation to simulate lattice-gas automata.
\newblock {\em Physical Review Letters}, 61(20):2332--2335, 1988.

\bibitem{chen-jun91}
Shiyi Chen, Hudong Chen, Gary~D. Doolen, Semion Gutman, and Minxu Lee.
\newblock A lattice gas model for thermohydrodynamics.
\newblock {\em Journal of Statistical Physics}, 62(5/6):1121--1151, 1991.

\bibitem{karniadakis-93}
George~Em Karniadakis and Steven~A. Orszag.
\newblock Nodes, modes and flow codes.
\newblock {\em Physics Today}, 46(3):34--42, 1993.

\bibitem{rothman-89}
Daniel~H. Rothman.
\newblock Negative-viscosity lattice gases.
\newblock {\em Journal of Statistical Physics}, 56(3/4):517--524, 1989.

\bibitem{chen-91}
S.~Chen, G.D. Doolean, K.~Eggert, D.~Grunau, and E.Y. Loh.
\newblock Local lattice-gas model for immiscible fluids.
\newblock {\em Physical Review A}, 43(12):7053--7056, 1991.

\bibitem{gustensen-91a}
Andrew~K. Gustensen and Daniel~H. Rothman.
\newblock A lattice-gas model for three immiscible fluids.
\newblock {\em Physica}, D(47):47--52, 1991.

\bibitem{gustensen-91b}
Andrew~K. Gustensen and Daniel~H. Rothman.
\newblock A galilean-invariant immiscible lattice gas.
\newblock {\em Physica}, D(47):53--63, 1991.

\bibitem{chen-sep89}
Hudong Chen, Shiyi Chen, Gary~D. Doolen, Y.C. Lee, and H.A. Rose.
\newblock Multithermodynamic phase lattice-gas automata incorporating
  interparticle potentials.
\newblock {\em Physical Review A}, 40(5):2850--2853, 1989.
\newblock Rapid Communications.

\bibitem{appert-90}
C\'ecile Appert and St\'ephane Zaleski.
\newblock Lattice gas with a liquid-gas transition.
\newblock {\em Physical Review Letters}, 64:1--4, 1990.

\bibitem{appert-91}
C\'ecile Appert, Daniel Rothman, and St\'ephane Zaleski.
\newblock A liquid-gas model on a lattice.
\newblock In Gary~D. Doolean, editor, {\em Lattice Gas Methods: Theory,
  Applications, and Hardware}, pages 85--96. Special Issues of Physica D,
  MIT/North Holland, 1991.

\bibitem{yepez-93}
Jeffrey Yepez.
\newblock A reversible lattice-gas with long-range interactions coupled to a
  heat bath.
\newblock In Gary~D. Doolean, editor, {\em Proceedings of the Pattern Formation
  and Lattice-Gas Automata Conference}. Fields Institute, American Mathematical
  Society, 1993.
\newblock To appear.

\bibitem{chen2}
Shiyi Chen, Hudong Chen, Daniel Mart\'inez, and William Matthaeus.
\newblock Lattice boltzmann model for simulation of magnetohydrodynamics.
\newblock {\em Physical Review Letters}, 67(27):3776--3779, 1991.

\bibitem{rothmann-90}
Daniel Rothman.
\newblock Macroscopic laws for immiscible two-phase flow in porous media:
  results from numerical experiments.
\newblock {\em Journal of Geophysics Research}, 95:8663, 1990.

\bibitem{chen-91b}
Shiyi Chen, Karen Diemer, Gary~D. Doolen, and Kenneth Eggert.
\newblock Lattice gas automata for flow through porous media.
\newblock In Gary~D. Doolean, editor, {\em Lattice Gas Methods: Theory,
  Applications, and Hardware}, pages 72--84. Special Issues of Physica D,
  MIT/North Holland, 1991.

\bibitem{margolus-93}
Norman Margolus.
\newblock Cam-8: a computer architecture based on cellular automata.
\newblock In Gary~D. Doolean, editor, {\em Proceedings of the Pattern Formation
  and Lattice-Gas Automata Conference}. Fields Institute, American Mathematical
  Society, 1993.
\newblock To appear.

\bibitem{boghosian-91}
Bruce~M. Boghosian.
\newblock Lattice gases illustrate the power of cellular automata in physics.
\newblock {\em Computers in Physics}, 5(6), Nov/Dec 1991.

\bibitem{hardy-76}
J.~Hardy, O.~de~Pazzis, and Y.~Pomeau.
\newblock Molecular dynamics of a classical lattice gas: Transport properties
  and time correlation functions.
\newblock {\em Physical Review A}, 13(5):1949--1961, 1976.

\bibitem{toffoli-77b}
Tommaso Toffoli.
\newblock Computation and construction universality of reversible cellular
  automata.
\newblock {\em Journal of Computer and System Sciences}, 15(2):213--231, 1977.

\bibitem{fredkin-82}
Edward Fredkin and Tommaso Toffoli.
\newblock Conservative logic.
\newblock {\em International Journal of Theoretical Physics}, 21(3/4):219--253,
  1982.

\bibitem{margolus-84}
Norman Margolus.
\newblock Physics-like models of computation.
\newblock {\em Physica}, 10D:81--95, 1984.

\bibitem{toffoli-84b}
Tommaso Toffoli.
\newblock Cam: A high-performance cellular-automaton machine.
\newblock {\em Physica}, 10D:195--204, 1984.
\newblock A demonstration TM-gas experiment was part of the CAMForth software
  distribution.

\bibitem{vichniac-84}
Gerard~Y. Vichniac.
\newblock Simulating physics with cellular automata.
\newblock {\em Physica}, 10D:96--116, 1984.

\bibitem{packard-85}
Norman~H. Packard and Stephen Wolfram.
\newblock Two-dimensional cellular automata.
\newblock {\em Journal of Statistical Physics}, 38(5/6):901--946, 1985.

\bibitem{Wolfram-85}
Stephen Wolfram and James~B. Salem.
\newblock Thermodynamics and hydrodynamics with cellular automata.
\newblock In Stephen Wolfram, editor, {\em Theory and Applications of Cellular
  Automata}, pages 362--365. World Scientific, 1986.
\newblock Submitted November 1985.

\bibitem{frisch-86}
Uriel Frisch, Brosl Hasslacher, and Yves Pomeau.
\newblock Lattice-gas automata for the navier-stokes equation.
\newblock {\em Physical Review Letters}, 56(14):1505--1508, 1986.

\bibitem{margolus-86}
Norman Margolus, Tommaso Toffoli, and G\'erand Vichniac.
\newblock Cellular-automata supercomputers for fluid-dynamics modeling.
\newblock {\em Physical Review Letters}, 56(16):1694--1696, 1986.

\bibitem{frisch-87}
Uriel Frisch, Dominique d'Humi\`eres, Brosl Hasslacher, Pierre Lallemand, Yves
  Pomeau, and Jean-Pierre Rivet.
\newblock Lattice gas hydrodynamics in two and three dimensions.
\newblock {\em Complex Systems}, 1:649--707, 1987.

\bibitem{henon-90}
Michel H\'enon.
\newblock Viscosity of a lattice gas.
\newblock In Gary~D. Doolean, editor, {\em Lattice Gas Methods for Partial
  Differential Equations}, pages 179--207. Santa Fe Institute, Addison-Wesley
  Publishing Company, 1990.

\bibitem{somers-89}
J.A. Somers and R.C. Rem.
\newblock The construction of efficient collision tables for fluid flow
  computations with cellular automata.
\newblock In P.~Manneville, N.~Boccara, G.Y. Vichniac, and R.~Bidaux, editors,
  {\em Cellular Automata and Modeling of Complex Physical Systems}, pages
  161--177. Springer-Verlag, Februrary 1989.
\newblock Proceedings of the Winter School, Les Houches, France.

\bibitem{margolus-90b}
Norman Margolus and Tommaso Toffoli.
\newblock Cellular automata machines.
\newblock In Gary~D. Doolean, editor, {\em Lattice Gas Methods for Partial
  Differential Equations}, pages 219--249. Santa Fe Institute, Addison-Wesley
  Publishing Company, 1990.
\newblock The first 8-module CAM-8 prototype was operational in the fall of
  1992.

\bibitem{margolus-90a}
Norman Margolus.
\newblock Parallel quantum computation.
\newblock In W.H. Zurek, editor, {\em Complexity, Entropy, and the Physics of
  Information,SFI Studies in the Sciences of Complexity, vol. VIII}, pages
  273--287. Addison-Wesley, 1990.

\bibitem{lloyd-93}
Seth Lloyd.
\newblock A technologically feasible quantum computer.
\newblock Complex System Group T-13, Los Alamos National Laborarory, Preprint.

\bibitem{biafore-94}
Michael Biafore.
\newblock Cellular automata for nanometer-scale computation.
\newblock {\em Physica D}, 70(3/4), 1994.

\bibitem{heitmann-93}
Detlef Heitmann and Jorg~P. Kotthaus.
\newblock The spectroscopy of quantum dot arrays.
\newblock {\em Physics Today}, 46(6):56--63, 1993.

\bibitem{bennett-79}
Charles~H. Bennett.
\newblock Logical reversibility of computation.
\newblock {\em IBM Journal of Research and Development}, 6:525--532, 1979.

\bibitem{bennett-82}
Charles~H. Bennett.
\newblock Thermodynamics of computation---a review.
\newblock {\em International Journal of Theoretical Physics}, 21:219--253,
  1982.

\bibitem{feynman-82}
Richard~P. Feynman.
\newblock Simulating physics with computers.
\newblock {\em International Journal of Theoretical Physics}, 21(6/7):467--488,
  1982.

\bibitem{alder-jun67}
B.J. Alder and T.E. Wainwright.
\newblock Velocity autocorrelations for hard spheres.
\newblock {\em Physical Review Letters}, 18(23):988--990, 1967.

\bibitem{pomeau-sep68}
Yves Pomeau.
\newblock A new kinetic theory for a dense classical gas.
\newblock {\em Physics Letters}, 27A(9):601--602, 1968.

\bibitem{ernst-nov70}
M.H. Ernst, E.H. Hauge, and J.M.M van Leeuwen.
\newblock Asymptotic time behavior of correlation functions.
\newblock {\em Physical Review Letters}, 25(18):1254--1256, 1970.

\bibitem{kirkpatrick-dec91}
T.R. Kirkpatrick and M.H. Ernst.
\newblock Kinetic theory for lattice-gas cellular automata.
\newblock {\em Physical Review A}, 44(12):8051--8061, 1991.

\bibitem{brito-dec91}
R.~Brito and M.H. Ernst.
\newblock Lattice gases in slab geometries.
\newblock {\em Physical Review A}, 44(12):8384--8687, 1991.

\bibitem{brito-jul92}
R.~Brito and M.H. Ernst.
\newblock Ring kinetic theory for tagged-particle problems in lattice gases.
\newblock {\em Physical Review A}, 46(2):875--887, 1992.

\bibitem{Li-jun93}
Chihwen Li and Chwan-Hwa Wu.
\newblock A particle-in-cell fluid model for radio frequency glow discharges.
\newblock {\em Computers in Physics}, 7(3):363--375, 1993.

\bibitem{teixeira-92}
Christopher Teixeira.
\newblock {\em Continuum Limit Of Lattice Gas Fluid Dynamics}.
\newblock PhD thesis, Massachusetts Institute of Technology, Department of
  Nuclear Engineering, 1992.
\newblock Kim Molvig Thesis Supervisor.

\bibitem{TMC-nov92}
TMC.
\newblock {\em CM5 Technical Summary}.
\newblock Thinking Machines Corporation, Cambridge, Massachusetts USA, 1992.

\bibitem{henon-92}
Michel H\'enon.
\newblock Implementation of the fchc lattice gas model on the connection
  machine.
\newblock {\em Journal of Statistical Physics}, 68(3/4):353--377, 1992.

\bibitem{normand-jul77}
Christiane Normand and Yves Pomeau.
\newblock Convective instability: A physicist's approach.
\newblock {\em Reviews of Modern Physics}, 49(3):581--624, 1977.

\bibitem{cohen-jan84}
E.G.D. Cohen.
\newblock The kinetic theory of fluids--an introduction.
\newblock {\em Physics Today}, pages 64--73, January 1984.

\bibitem{careri-84}
Giorgio Careri.
\newblock {\em Order And Disorder In Matter}.
\newblock The Benjamin/Cummings Publishing Company, 1984.

\bibitem{baker-90}
Gregory~L. Baker and Jerry~P. Gollub.
\newblock {\em Chaotic Dynamics an introduction}.
\newblock Cambridge University Press, 1990.

\bibitem{burges-87}
Christopher Burges and St\'ephane Zaleski.
\newblock Buoyant mixtures of cellular automaton gases.
\newblock {\em Complex Systems}, 1:31--50, 1987.

\bibitem{ernst-jsp66-92}
M.H. Ernst and Shankar~P. Das.
\newblock Thermal cellular automata fluids.
\newblock {\em Journal of Statistical Physics}, 66(1/2):465--483, 1992.

\bibitem{grosfils-jsp-93}
P.~Grosfils, J.P. Boon, and P.~Lallemand.
\newblock 19-bit thermal lattice-gas automaton.
\newblock {\em Journal of Statistical Physics}, 1993.
\newblock To appear.

\bibitem{Shimomura-90}
Tsutomu Shimomura, Gary~D. Doolen, Brosl Hasslacher, and Castor Fu.
\newblock Calculations using lattice gas techniques.
\newblock In Gary~D. Doolean, editor, {\em Lattice Gas Methods for Partial
  Differential Equations}, pages 3--9. Santa Fe Institute, Addison-Wesley
  Publishing Company, 1990.

\bibitem{schen-90}
S.~Chen, G.D. Doolean, K.~Eggert, D.~Grunau, and E.Y. Loh.
\newblock Lattice gas simulations of one and two-phase fluid flows using the
  connection machine-2.
\newblock In A.S. Alves, editor, {\em Discrete Models of Fluid Dynamics}, pages
  232--248. World Scientific, September 1990.
\newblock Series on Advances in Mathematics for Applied Sciences, Vol. 2.

\bibitem{wolfram-86}
Stephen Wolfram.
\newblock Cellular automaton fluids 1: Basic theory.
\newblock {\em Journal of Statistical Physics}, 45(3/4):471--526, 1986.

\bibitem{TMC-nov90}
TMC.
\newblock {\em C* Programming Guide}.
\newblock Thinking Machines Corporation, Cambridge, Massachusetts USA, 1990.
\newblock Version 6.0.

\end{thebibliography}
\end{document}